\documentclass[12pt, doublespace]{article}

\usepackage{amsthm}
\usepackage{amsmath}
\usepackage{amssymb}
\usepackage{bm}
\usepackage{enumerate}
\usepackage{epsf}
\usepackage{graphicx}
\usepackage[mathscr]{eucal}
\usepackage{pstricks,pst-node,pst-text,pst-3d}
\usepackage{titlesec}

\newtheorem{definition}{Definition}
\newtheorem{theorem}{Theorem}
\newtheorem{remark}{Remark}

\titleformat{\section}[hang]{\normalfont\filcenter}{\thesection.}{0.3cm}{}

\def\cqd{\hfill \rule{2.25mm}{2.25mm}\vspace{10pt}}

\begin{document}
\title{\bf\Large Minimum $\phi $-divergence estimation in constrained latent class models for binary data}
\author{}
\date{}
\maketitle

\thispagestyle{empty}

\section*{{\bf Abstract}}

The main purpose of this paper is to introduce and study the behavior of minimum $\phi $-divergence estimators as an alternative to the maximum likelihood estimator in latent class models for binary items. As it will become clear below, minimum $\phi $-divergence estimators are a natural extension of the maximum likelihood estimator. The asymptotic properties of minimum $\phi $-divergence estimators for latent class models for binary data are developed. Finally, to compare the efficiency and robustness of these new estimators with those obtained through maximum likelihood when the sample size is not big enough to apply the asymptotic results, we have carried out a simulation study.


Keywords: Latent class models, Minimum phi-divergence estimator,  Maximum likelihood estimator, Asymptotic distribution.

\newpage

\section{{\bf Introduction}}

Latent class models (LCM) were introduced in Lazarsfeld (1950) as a tool for studying categorical data analysis. Since then, many papers have been published with applications of LCM in different areas; see e.g. Hagenaars \& Cutcheon (2002), Langeheine \& Rost (1988), Rost \& Langeheine (1997), Collins \& Lanza (2010), Biemer (2011); LCM models are specially important in behavioral and social sciences (e.g. Hagenaars \& Cutcheon (2002), Abar \& Loken (2010), Caldwell et al. (2009), Coffman et al. (2007), Feldman et~al. (2009), Gerber et~al. (2009), Laska et~al. (2009), Nylund et~al. (2007)).

In order to be self-contained and to fix notation, let us introduce the problem we deal with in this paper. We shall formulate the problem of LCM for binary data in the same way as it appears in  Formann (1985), and we shall use the parametrization considered in that paper. Consider a set ${\cal P}$ of $N$ people, ${\cal P}:=\{ P_1, ..., P_N\} $. Each person $P_{v}$ answers to $k$ dichotomous items $I_1, ..., I_k;$ let us denote by $y_{v i}$ the answer of person $P_{v}$ to item $I_i, i=1, ..., k,$ i.e.

$$ y_{v i}:=\left\{ \begin{array}{cl}1 & \mbox{~if~the~answer~of~} P_{v} \mbox{~to~} I_i \mbox{~is~correct} \\ 0 & \mbox{~otherwise} \end{array} \right. .$$

Let ${\bf y}_{v}:=(y_{v1}, ..., y_{vk})$ denote a generic pattern of right and wrong answers to the $k$ items given by $P_{v}.$ To explain the statistical relationships of the observed variables, a categorical latent variable (categorical unobservable variable) is postulated to exist, whose different levels partition set ${\cal P}$ into $m$ mutually exclusive and exhaustive latent classes. Let us denote these classes by $C_1, ..., C_m$ and their corresponding relative sizes by $w_1, ..., w_m;$ thus, $w_j$ denotes the probability of a randomly selected person $P_{v} \in {\cal P}$ belongs to class $C_j,$ i.e.

$$ w_j=Pr(P_{v}\in C_j),\, j=1, ..., m.$$

We denote by $p_{ji}$ the probability of a right answer of $P_{v}$ to the item $I_i$ under the assumption that $P_{v}$ is in class $C_j:$

$$ p_{ji}= Pr(y_{v i}=1 | P_{v}\in C_j).$$

Let ${\bf y_{\nu }}$ be a possible answer vector. We shall assume that in each class the answers for the different questions are stochastically independent; therefore, we can write

$$ Pr({\bf y_{\nu }}| P_{v}\in C_j) = \prod_{i=1}^k p_{ji}^{y_{\nu i}} (1-p_{ji})^{1-y_{\nu i}},$$
and
\begin{equation}\label{eq1}
Pr({\bf y_{\nu }}) = \sum_{j=1}^m w_j \prod_{i=1}^k p_{ji}^{y_{\nu i}} (1-p_{ji})^{1-y_{\nu i}}.
\end{equation}

There are $2^k$ possible answer vectors ${\bf y_{\nu }}$ whose probability of occurrence are given by \eqref{eq1}; they constitute the manifest probabilities for the items $I_1, ..., I_k$ in the population given by $P_1, ..., P_N.$ We will denote by $N_{\nu },\, \nu =1, ..., 2^k,$ the number of times that the sequence ${\bf y_{\nu }}$ appears in an $N$-sample and

$$ {\bf \hat{p}}=(N_1/N, ..., N_{2^k}/N).$$

The likelihood function $L$ is given by

\begin{equation}\label{eq2}
L(w_1, ..., w_m, p_{11}, ..., p_{mk})=Pr (N_1=n_1, ..., N_{2^k}=n_{2^k})= {N!\over \prod_{\nu =1}^{2^k} n_{\nu }!} \prod_{\nu =1}^{2^k} Pr ({\bf y_{\nu }})^{n_{\nu }}.
\end{equation}

By $n_{\nu }$ we are denoting a realization of the random variable $N_{\nu }, \nu =1, ..., 2^k.$ In this model the unknown parameters are $w_j, j=1, ..., m$ and $p_{ji}, j=1, ..., m, i=1, ..., k.$ These parameters can be estimated using the maximum likelihood estimator (e.g. McHugh (1956), Lazarsfeld \& Henry (1968), Clogg (1995)). In order to avoid the problem of obtaining uninterpretable estimates for the item latent probabilities lying outside the interval $[0,1],$ some authors (Lazarsfeld \& Henry (1968), Formann (1976), Formann (1977), Formann (1978), Formann (1982), Formann (1985)) proposed a linear-logistic parametrization for the probabilities $w_j$ and $p_{ji}$ given by

$$
p_{ji}={exp (x_{ji})\over 1 +exp (x_{ji})},\, \, j=1, ..., m,\, \, i=1, ..., k, $$ and $$ w_j={exp (z_j)\over {\displaystyle \sum_{h=1}^m exp (z_h)}},\, \, j=1, ..., m.
$$

Next, restrictions are introduced relating these parameters to some explanatory variables, parameters $\lambda_r, r=1, ..., t$ and $\eta_s, s=1, ..., u$, so the final model is given by

\begin{equation}\label{F1} p_{ji}={exp ({\displaystyle \sum_{r=1}^t q_{jir} \lambda_r + c_{ji}})\over 1 +exp ({\displaystyle \sum_{r=1}^t q_{jir} \lambda_r + c_{ji}})},\, \, j=1, ..., m,\, \, i=1, ..., k,
\end{equation}
and

\begin{equation}\label{F2} w_j={exp ({\displaystyle \sum_{r=1}^u v_{jr} \eta_r + d_{j}})\over {\displaystyle \sum_{h=1}^m exp (\sum_{r=1}^u v_{hr} \eta_r + d_{h})}},\, \, j=1, ..., m, \end{equation}
where

$$ {\bf Q}_r=(q_{jir})_{\stackrel{j=1, ..., m}{i=1, ..., k}}, r=1, ..., t,\, \, {\bf C}=(c_{ji})_{\stackrel{j=1, ..., m}{i=1, ..., k}},\, \, {\bf V}=(v_{jr})_{\stackrel{j=1, ..., m}{r=1, ..., u}},\, \, {\bf d}=(d_{j})_{j=1, ..., m}, $$ are fixed. Matrix ${\bf Q}$ specifies to what extent the predictors defined through parameters $\lambda_r$ are relevant for each $x_{ji}.$ The terms $c_{ji}$ were introduced to include the possibility that certain $p_{ji}$ are fixed to certain previously determined values; this possibility was considered by Goodman (1974). The same applies for matrix ${\bf V}$: thus, ${\bf V}$ specifies to which amount $\eta_s$ is relevant for each $z_j.$ The terms $d_j$ are introduced to include the possibility that certain $z_j$ are fixed to certain previously determined values.

Consequently, in this case the vector of unknown parameters $\bm \theta $ in the LCM for binary data is given by $\bm \theta := (\bm \lambda , \bm \eta ),$ where $\bm \lambda $ and $\bm \eta $ are defined as $ \bm \lambda := (\lambda_1, ..., \lambda_t),\, \bm \eta := (\eta_1, ..., \eta_u).$

By $\bm \Theta $ we shall denote the set in which the parameter $\bm \theta $ varies, i.e. the parametric space. Thus, we have $t+u$ unknown parameters that can be estimated by maximum likelihood using \eqref{eq2}, and from these estimations, the corresponding estimations of $p_{ji}, w_j, j=1, ..., m, i=1, ..., k.$ In the following we shall assume this parametrization.

The main purpose of this paper is to present a new procedure for estimating $ p_{ji}, w_j$ through previous estimations of the parameters $\lambda_i, i=1, ..., t$ and $\eta_j, j=1,..., u.$ To achieve this task, we will introduce in the context of LCM for binary data a new family of estimators based on divergence measures, the so-called minimum $\phi $-divergence estimators. This family of estimators contains as a particular case the maximum likelihood estimator. Minimum $\phi $-divergence estimators were first introduced in Morales et al. (1995) and since then, many interesting estimation problems have been solved using minimum $\phi $-divergence estimators, see e.g. Pardo (2006), where it is pointed out the importance of some minimum $\phi $-divergence estimators for different statistical problems. In the same way, in this paper we establish the importance of these estimators for estimating the parameters in the LCM for binary data.

The rest of the paper is organized as follows: In Section 2 we introduce the definition of minimum $\phi $-divergence estimator in the context of LCM for binary data. Its asymptotic behavior is presented in Section 3, as well as its asymptotic distribution. Section 4 presents a case study based on an example proposed by Coleman (1964) and studied later by Goodman (1974). The behavior of minimum $\phi $-divergence estimators when the sample size is not big enough to apply the asymptotical results of Section 3 is carried out in Section 5 on the basis of a simulation study.  Last section is devoted to the conclusions. Finally, in an appendix we provide the proofs of the results presented in the paper.

\section{{\bf Minimum $\phi $-divergence estimator in LCM}}

In this section we are going to introduce the minimum $\phi $-divergencce estimator as a natural extension of the maximum likelihood estimator. In the following, we denote $Pr({\bf y_{\nu }})$ by $p({\bf y_{\nu }}, \bm \lambda , \bm \eta ).$ Based on \eqref{eq2}, the maximum likelihood estimator is obtained by maximizing in $\bm \lambda $ and $\bm \eta $ the log-likelihood function,

\begin{equation}\label{eq3}
\sum_{\nu =1}^{2^k} N_{\nu } \log p({\bf y_{\nu }}, \bm \lambda , \bm \eta ).
\end{equation}

The expression \eqref{eq3} can be written as

\begin{eqnarray*}
\sum_{\nu =1}^{2^k} N_{\nu } \log p({\bf y_{\nu }}, \bm \lambda , \bm \eta ) & = & N \sum_{\nu =1}^{2^k} {N_{\nu }\over N} \log p({\bf y_{\nu }}, \bm \lambda , \bm \eta ) \\
& = & -N \sum_{\nu =1}^{2^k} \hat{p}_{\nu } \log {1\over p({\bf y_{\nu }}, \bm \lambda , \bm \eta )} - N \sum_{\nu =1}^{2^k} \hat{p}_{\nu } \log \hat{p}_{\nu } + N \sum_{\nu =1}^{2^k} \hat{p}_{\nu } \log \hat{p}_{\nu } \\
& = & -N \sum_{\nu =1}^{2^k} \hat{p}_{\nu } \log {\hat{p}_{\nu }\over p({\bf y_{\nu }}, \bm \lambda , \bm \eta )} + N \sum_{\nu =1}^{2^k} \hat{p}_{\nu } \log \hat{p}_{\nu } \\
& = & - N D_{Kullback} ({\bf \hat{p}}, {\bf p}(\bm \lambda , \bm \eta )) + constant,
\end{eqnarray*}
being $ {\bf p}(\bm \lambda , \bm \eta )) = (p(y_1, \bm \lambda , \bm \eta ), ..., p(y_{2^k}, \bm \lambda , \bm \eta )),$ and

\begin{equation}\label{eq4}
D_{Kullback} ({\bf \hat{p}}, {\bf p}(\bm \lambda , \bm \eta )) = \sum_{\nu =1}^{2^k} \hat{p}_{\nu } \log {\hat{p}_{\nu }\over p({\bf y_{\nu }}, \bm \lambda , \bm \eta )},
\end{equation}
the Kullback-Leibler divergence measure between the probability vectors ${\bf \hat{p}}$ and ${\bf p}(\bm \lambda , \bm \eta ).$ Therefore, the problem of maximizing \eqref{eq3} in $\bm \lambda $ and $\bm \eta $ is equivalent to the problem of minimizing \eqref{eq4} in $\bm \lambda $ and $\bm \eta $. Consequently, the value $\hat{\bm \theta }= (\hat{\bm \lambda }, \hat{\bm \eta })$ that minimizes $\bm \theta = (\bm \lambda , \bm \eta )$ in the Kullback-Leibler divergence is the maximum likelihood estimator of the parameters for the LCM for binary data or, in other words, the {\it minimum Kullback-Leibler divergence estimator}. We shall denote it by $\hat{\bm \theta }$ or

$$ \hat{\bm \theta }_{Kullback} := arg \min_{(\bm \lambda , \bm \eta )\in \bm \Theta } D_{Kullback} ({\bf \hat{p}}, {\bf p}(\bm \lambda , \bm \eta )).$$

Thus, we can observe that the maximum likelihood estimator for the LCM for binary data consists of the values $\hat{\bm \lambda }$ and $\hat{\bm \eta }$ for which the Kullback-Leibler divergence measure between the probability vectors ${\bf \hat{p}}$ and ${\bf p}(\hat{\bm \lambda }, \hat{\bm \eta })$ minimizes. If $D$ is a measure of distance between ${\bf \hat{p}}$ and ${\bf p}(\bm \lambda , \bm \eta ),$ we can generalize the concept of maximum likelihood estimator (or minimum Kullback-Leibler divergence estimator) by

$$ \hat{\bm \theta }_{D} := arg \min_{(\bm \lambda , \bm \eta )\in \bm \Theta } D({\bf \hat{p}}, {\bf p}(\bm \lambda , \bm \eta )).$$

A question now arises: which measures of distance are suitable for generalizing the Kullback-Leibler divergence? To answer this question, it is necessary to keep in mind that the Kullback-Leibler divergence measure between ${\bf \hat{p}}$ and ${\bf p}(\bm \lambda , \bm \eta )$ is a particular case of the family of $\phi ${\bf -divergence measures} introduced in Csisz\'ar (1967) through

\begin{equation}\label{eq5}
D_{\phi }({\bf \hat{p}}, {\bf p}(\bm \lambda , \bm \eta ))= \sum_{\nu =1}^{2^k} p({\bf y_{\nu }}, \bm \lambda , \bm \eta ) \phi \left( {\hat{p}_{\nu }\over p({\bf y_{\nu }}, \bm \lambda , \bm \eta )}\right) ,
\end{equation}
where $\phi $ is a convex function for $x>0$ satisfying $\phi (1)=0, 0\phi (0/0)=0$ and

$$ 0\phi (p/0)=p\lim_{x\rightarrow \infty } {\phi (x)\over x}.$$
Let us denote the set of all functions $\phi $ in these conditions by $\Phi^*$. In particular, taking $\phi_0 (x)=x\log x-x+1,$ we obtain the Kullback-Leibler divergence measure, i.e.

$$ D_{\phi_0 }({\bf \hat{p}}, {\bf p}(\bm \lambda , \bm \eta ))= D_{Kullback}({\bf \hat{p}}, {\bf p}(\bm \lambda , \bm \eta )).$$

Now, let $\phi \in \Phi^*$ be differentiable at $x=1;$ then, the function $ \psi (x):= \phi (x) - \phi '(1)(x-1)$ also belongs to $\Phi^*$ and has the additional property that $\psi '(1)=0.$ This property, together with the convexity, implies that $\psi (x)\geq 0$ for any $x\geq 0.$ Moreover,

\begin{eqnarray*}
D_{\psi }({\bf \hat{p}}, {\bf p}(\bm \lambda , \bm \eta )) & = & \sum_{\nu =1}^{2^k} p({\bf y_{\nu }}, \bm \lambda , \bm \eta ) \psi \left( {\hat{p}_{\nu }\over p({\bf y_{\nu }}, \bm \lambda , \bm \eta )}\right)  \\
 & = & \sum_{\nu =1}^{2^k} p({\bf y_{\nu }}, \bm \lambda , \bm \eta ) \left( \phi \left( {\hat{p}_{\nu }\over p({\bf y_{\nu }}, \bm \lambda , \bm \eta )}\right) - \phi'(1) \left[ {\hat{p}_{\nu }\over p({\bf y_{\nu }}, \bm \lambda , \bm \eta )}-1\right] \right) \\
 & = & \sum_{\nu =1}^{2^k} p({\bf y_{\nu }}, \bm \lambda , \bm \eta ) \phi \left( {\hat{p}_{\nu }\over p({\bf y_{\nu }}, \bm \lambda , \bm \eta )}\right) \\
 & = & D_{\phi }({\bf \hat{p}}, {\bf p}(\bm \lambda , \bm \eta ))
\end{eqnarray*}

Since the two divergence measures coincide, we can consider the set $\Phi^*$ to be equivalent to the set $ \Phi := \Phi^* \cap \{ \phi: \phi '(1)=0\} .$ More details about $\phi $-divergence measures can be seen in Cressie and Pardo (2002) and Pardo (2006).

Based on the previous results, we can define the {\bf minimum $\phi $-divergence estimator} (M$\phi $E) in LCM for binary data in the following way:

\begin{definition}
Given a LCM for binary data with parameters $\bm \lambda =(\lambda_1, ..., \lambda_t )$ and $\bm \eta =(\eta_1, ..., \eta_u ),$ the M$\phi $E of $\bm \theta = (\bm \lambda , \bm \eta )$ is any $\hat{\bm \theta}_{\phi }$ satisfying

$$  \hat{\bm \theta}_{\phi } = arg \min_{(\bm \lambda , \bm \eta )\in \bm \Theta } D_{\phi } ({\bf \hat{p}}, {\bf p}(\bm \lambda , \bm \eta )).$$
\end{definition}

\begin{remark}
From a practical point of view, we must solve the following system of equations:

$$ {\partial D_{\phi } ({\bf \hat{p}}, {\bf p}(\bm \lambda , \bm \eta ))\over \partial s_j}= 0,\, j=1, ..., u+t $$ being

\begin{equation}\label{eq8}
s_j:= \left\{ \begin{array}{ll} \lambda_j, & j =1, ..., t \\ \eta_{j-t}, & j =t+1, ..., t+u\end{array}\right. .
\end{equation}

It is not difficult to see that

$$ {\partial D_{\phi } ({\bf \hat{p}}, {\bf p}(\bm \lambda , \bm \eta ))\over \partial \lambda_{\alpha }}= \sum_{\nu =1}^{2^k} \left\{ {\partial p({\bf y_{\nu }}, \bm \lambda , \bm \eta )\over \partial \lambda_{\alpha }} \phi \left( {\hat{p}_{\nu }\over p({\bf y_{\nu }}, \bm \lambda , \bm \eta )} \right) - {\hat{p}_{\nu }\over p({\bf y_{\nu }}, \bm \lambda , \bm \eta )} \phi '\left( {\hat{p}_{\nu }\over p({\bf y_{\nu }}, \bm \lambda , \bm \eta )} \right) {\partial p({\bf y_{\nu }}, \bm \lambda , \bm \eta )\over \partial \lambda_{\alpha }}  \right\} .$$

$$ {\partial D_{\phi } ({\bf \hat{p}}, {\bf p}(\bm \lambda , \bm \eta ))\over \partial \eta_{\beta }}= \sum_{\nu =1}^{2^k} \left\{ {\partial p({\bf y_{\nu }}, \bm \lambda , \bm \eta )\over \partial \eta_{\beta }} \phi \left( {\hat{p}_{\nu }\over p({\bf y_{\nu }}, \bm \lambda , \bm \eta )} \right) - {\hat{p}_{\nu }\over p({\bf y_{\nu }}, \bm \lambda , \bm \eta )} \phi '\left( {\hat{p}_{\nu }\over p({\bf y_{\nu }}, \bm \lambda , \bm \eta )} \right) {\partial p({\bf y_{\nu }}, \bm \lambda , \bm \eta )\over \partial \eta_{\beta }}  \right\} ,$$
where

$$ {\partial p({\bf y_{\nu }}, \bm \lambda , \bm \eta )\over \partial \lambda_{\alpha }} = \sum_{j=1}^m w_j Pr ({\bf y_{\nu }} | P_{\nu }\in C_j ) \sum_{i=1}^k q_{ji\alpha } (y_{\nu i}- p_{ji}), \, \, \alpha =1, ..., t. $$

$$ {\partial p({\bf y_{\nu }}, \bm \lambda , \bm \eta )\over \partial \eta_{\beta }} = \sum_{j=1}^m w_j Pr ({\bf y_{\nu }} | P_{\nu }\in C_j )\left[ v_{j\beta } - \sum_{h=1}^m w_h v_{h\beta } \right] ,\, \, \beta =1, ..., u.$$

The calculus for obtaining these derivatives can be seen in the appendix.
\end{remark}

The previous equations constitute the necessary conditions for $D_{\phi } ({\bf \hat{p}}, {\bf p}(\bm \lambda , \bm \eta ))$ to have an extreme point at $\bm \theta^* = (\lambda_1^*, ...., \lambda_t^*, \eta_1^*, ..., \eta_u^*)$, but in general it is difficult to check for the mixture multinomial model whether it is indeed a minimum phi-divergence estimator. Apart from the problem that a solution of the previous system may fail to minimize $D_{\phi } ({\bf \hat{p}}, {\bf p}(\bm \lambda , \bm \eta )),$ we have to deal with the problem that several minimums of $D_{\phi } ({\bf \hat{p}}, {\bf p}(\bm \lambda , \bm \eta ))$ could exist. In order to obtain a good approximation of a global minimum and avoid a local minimum or a stationary point, we present in Section 5 a multistart optimization algorithm. It is noteworthy that these problems also appear when dealing with MLE (see Formann (1992) for more details).

In next section we study the behavior of the M$\phi $E in LCM for binary data for large sample sizes, i.e. the asymptotic behavior of the M$\phi $E.

\section{{\bf Asymptotic properties of the M$\phi $E in LCM for binary data}}

Let us denote by $(\bm \lambda_0, \bm \eta_0)=(\lambda_1^0, ..., \lambda_t^0, \eta_1^0, ..., \eta_u^0)$ the true value of the parameter $(\bm \lambda , \bm \eta )$ and let us assume that it is an interior point of the parameter space $\bm \Theta $. Let us denote by $\bm \Delta_{2^k}$ the set

$$ \bm \Delta_{2^k} := \left\{ {\bf p}=(p_1, ..., p_{2^k})^T : p_{\nu }\geq 0,\, \nu =1, ..., 2^k,\, \sum_{\nu =1}^{2^k} p_{\nu }=1\right\} .$$ In this section we shall assume that  Birch's conditions hold:

\begin{enumerate}[i)]
\item $p ({\bf y_{\nu }}, \bm \lambda_0, \bm \eta_0 )>0,\, \nu =1, ..., 2^k.$ Thus,
$${\bf p} (\bm \lambda_0, \bm \eta_0)=(p (y_{1}, \bm \lambda_0, \bm \eta_0), ..., p (y_{2^k}, \bm \lambda_0, \bm \eta_0))$$ is an interior point of $ \bm \Delta_{2^k}.$ In the following, and in order to avoid hard notation, we will denote $p ({\bf y_{\nu }}, \bm \lambda_0, \bm \eta_0 )$ by $p_{\nu }(\bm \lambda_0, \bm \eta_0).$
\item The mapping ${\bf p}:\bm \Theta \rightarrow \bm \Delta_{2^k}$ assigning to any $(\bm \lambda , \bm \eta )$ the vector ${\bf p} (\bm \lambda , \bm \eta )$ is continuous and totally differentiable at $(\bm \lambda_0 , \bm \eta_0).$
\item The Jacobian matrix

$$ {\bf J}(\bm \lambda_0, \bm \eta_0 ):= \left( {\partial p_{\nu }(\bm \lambda_0, \bm \eta_0 )\over \partial s_j }\right)_{\stackrel{ \nu =1, ..., 2^k}{j=1,..., t+u}} $$ is of rank $t+u.$

%

\item The inverse mapping of ${\bf p}$ is continuous at ${\bf p}(\bm \lambda_0, \bm \eta_0 ).$
\end{enumerate}

Now, the following can be proved:

\begin{theorem}\label{teo1}
Suppose $\phi (t)$ is twice continuously differentiable at any $t>0.$ Under Birch's conditions, the M$\phi $E, $\hat{\bm \theta }_{\phi },$ for the LCM for binary data satisfies

$$ \hat{\bm \theta }_{\phi }= (\bm \lambda_0, \bm \eta_0 )^T+ ({\bf A}(\bm \lambda_0, \bm \eta_0 )^T{\bf A}(\bm \lambda_0, \bm \eta_0 ))^{-1} {\bf A}(\bm \lambda_0, \bm \eta_0 )^T {\bf D}_{{\bf p}(\bm \lambda_0, \bm \eta_0 )}^{-{1\over 2}} (\hat{\bf p} - {\bf p}(\bm \lambda_0, \bm \eta_0 )) + o(\| \hat{\bf p} - {\bf p}(\bm \lambda_0, \bm \eta_0 )\| ),$$ where
$ {\bf A}(\bm \lambda_0, \bm \eta_0 ):={\bf D}_{{\bf p}(\bm \lambda_0, \bm \eta_0 )}^{-{1\over 2}} {\bf J}(\bm \lambda_0, \bm \eta_0 )$ and by ${\bf D}_{{\bf p}(\bm \lambda_0, \bm \eta_0 )}$ we are denoting the diagonal matrix whose diagonal is given by ${\bf p}(\bm \lambda_0, \bm \eta_0 )$.
\end{theorem}

{\bf Proof:} See Appendix.

We can observe in this theorem that the expansion obtained for the M$\phi $E in LCM for binary data does not depend on the function $\phi .$ This fact is very important because, based on it, in next theorem we shall establish that the asymptotic distribution of $\hat{\bm \theta }_{\phi }$ does not depend on $\phi .$

\begin{theorem}\label{teo2}
Under the assumptions of the previous theorem, the M$\phi $E, $\hat{\bm \theta }_{\phi },$ for the LCM for binary data satisfies

$$ \sqrt{N}( \hat{\bm \theta}_{\phi}- (\bm \lambda_0, \bm \eta_0 )^T) {\overset{L}{\underset{N\rightarrow \infty }{\longrightarrow }}} {\cal N}({\bf 0} , ({\bf A}(\bm \lambda_0, \bm \eta_0 )^T{\bf A}(\bm \lambda_0, \bm \eta_0 ))^{-1}).$$

\end{theorem}

{\bf Proof:} See Appendix.

If we pay attention to the asymptotic variance-covariance matrix of the M$\phi $E, we can see that this matrix is the inverse of the Fisher information matrix of the model under consideration. Therefore, the M$\phi $E are BAN (Best Asymptotically Normal) estimators and their efficiency coincides with the efficiency of the MLE for big sample sizes, i.e. from an asymptotic point of view.

On the other hand, we usually have to work with samples whose size is not big enough to apply the previous results, and in this case the behavior of the M$\phi $E may be different for different functions $\phi $. We shall study the behavior of $\hat{\bm \theta}_{\phi}$ in this case on the basis of a simulation study in Section 5 in order to clarify this point.

Finally, let us present a result in relation to the estimated manifest probabilities, ${\bf p}(\hat{\bm \theta }_{\phi }).$

\begin{theorem}\label{teo3}
Under the assumptions of the previous theorems, the estimated manifest probabilities satisfy

$$ \sqrt{N}( {\bf p}(\hat{\bm \theta}_{\phi})- {\bf p}(\bm \lambda_0, \bm \eta_0 )) {\overset{L}{\underset{N\rightarrow \infty }{\longrightarrow }}} {\cal N}({\bf 0} ,  {\bf J}(\bm \lambda_0 , \bm \eta_0)^T \left( {\bf A}(\bm \lambda_0, \bm \eta_0 )^t {\bf A}(\bm \lambda_0, \bm \eta_0 ) \right)^{-1} {\bf J}(\bm \lambda_0 , \bm \eta_0)).$$
\end{theorem}

{\bf Proof:} See Appendix.

\section{{\bf A numerical example}}

In order to study the M$\phi $E proposed in this paper we have considered the interview data collected by Coleman (1964) and analized later in Goodman (1974); this model is explained in Formann (1982) and Formann (1985). The experiment consists in evaluating the answers of 6658 schoolboys to two questions about their membership in the ``leading crowd" on two occasions $t_1$ and $t_2$ (October, 1957 and May, 1958). Thus, in this model we have 4 questions and there are four manifest variables (answers to both questions at both moments); these answers can only be ``low" (value 0) and ``high" (value 1), so that the manifest variables are dichotomous. The sample data is given in next table:

\begin{center}
\begin{tabular}{c|cccc}
October, 1957/ May, 1958 & 00 & 01 & 10 & 11 \\
\hline 00                & 1090 & 641 & 172 & 159 \\
01                       & 602  &1299 & 115 & 313 \\
10                       & 133  & 86  & 192 & 183 \\
11                       & 81   & 217 & 233 & 942 \\
\end{tabular}
\end{center}

Next, 4 latent classes are considered, namely

$C_1\equiv $ low agreement in question 1 and low agreement in question 2.

$C_2\equiv $ low agreement in question 1 and high agreement in question 2.

$C_3\equiv $ high agreement in question 1 and low agreement in question 2.

$C_4\equiv $ high agreement in question 1 and high agreement in question 2.

There are 16 probability values $p_{ji}$ to be estimated; we consider the first hypothesis appearing in Formann (1985), namely ``The attitudinal changes between times $t_1$ and $t_2$ are dependent on the positions (low, high) of the respective classes on the underlying attitudinal scales at $t_1$". Thus, a model with 8 parameters $\lambda_i$ is considered; $\lambda_1$ means low agreement in the first question at time $t_1$, $\lambda_2$ means high agreement in the first question at time $t_1$, $\lambda_3$ means low agreement in the second question at time $t_1$, $\lambda_4$ means high agreement in the second question at time $t_1$, and $\lambda_5, \lambda_6, \lambda_7, \lambda_8$ are the same parameters at time $t_2.$ We write the values for matrices ${\bf Q}_i$ as they appear in Formann (1985). In our notation, the matrices ${\bf Q}_i$ can be derived considering the $i$-th column in the table and dividing it in four columns of four elements each (each corresponding to a latent class).

\begin{center}
\begin{tabular}{c|c|cccccccc}
Class & Item & $\lambda_1$ &  $\lambda_2$ & $\lambda_3$ & $\lambda_4$ & $\lambda_5$ & $\lambda_6$ & $\lambda_7$ & $\lambda_8$ \\ \hline 1 & 1 & 1 & 0 & 0 & 0 & 0 & 0 & 0 & 0 \\ & 2 & 0 & 0 & 1 & 0 & 0 & 0 & 0 & 0 \\ & 3 & 0 & 0 & 0 & 0 & 1 & 0 & 0 & 0 \\ & 4 & 0 & 0 & 0 & 0 & 0 & 0 & 1 & 0 \\ \hline 2 & 1 & 1 & 0 & 0 & 0 & 0 & 0 & 0 & 0 \\ & 2 & 0 & 0 & 0 & 1 & 0 & 0 & 0 & 0 \\ & 3 & 0 & 0 & 0 & 0 & 1 & 0 & 0 & 0 \\ & 4 & 0 & 0 & 0 & 0 & 0 & 0 & 0 & 1 \\ \hline 3 & 1 & 0 & 1 & 0 & 0 & 0 & 0 & 0 & 0 \\ & 2 & 0 & 0 & 1 & 0 & 0 & 0 & 0 & 0 \\ & 3 & 0 & 0 & 0 & 0 & 0 & 1 & 0 & 0 \\ & 4 & 0 & 0 & 0 & 0 & 0 & 0 & 1 & 0 \\ \hline 4 & 1 & 0 & 1 & 0 & 0 & 0 & 0 & 0 & 0 \\ & 2 & 0 & 0 & 0 & 1 & 0 & 0 & 0 & 0 \\ & 3 & 0 & 0 & 0 & 0 & 0 & 1 & 0 & 0 \\ & 4 & 0 & 0 & 0 & 0 & 0 & 0 & 0 & 1 \\ \hline
\end{tabular}
\end{center}

Note that the hypothesis is that the attitudinal changes between times $t_1$ and $t_2$ are dependent upon the items as well as on the classes. For this reason, the part corresponding each latent class can be partitioned in four submatrices of size 2$\times $4. The submatrices lying on the main diagonal are the same by the hypothesis defining the model and the two other submatrices are null. The differences among them are due to the differences in the latent classes. Next, $c_{ij}=0,\, \, \forall i, j$ (as we have explained when values $c_{ij}$ were introduced in Section 1). Finally, 4 parameters $\eta_j$ are considered, taking as matrix ${\mathbf V}$ the identity matrix and $d_j=0,\, \forall j.$


It is noteworthy that our model assumes that answers to the questions are
conditionally independent given the latent class. In this example, we are dealing with repeated
responses to two questions, so this assumption may be unrealistic. However, this assumption is made in the original paper of Goodman and we follow this assumption for the sake of the example.

In order to get the estimations of the parameters, we shall consider in our study the family of $\phi $-divergences introduced in Cressie and Read (1984). This family of $\phi $-divergence measures, called the {\it power-divergence family}, is obtained from \eqref{eq5} with

\begin{equation}\label{eq9}
\phi (x)\equiv \phi_a(x)=\left\{ \begin{array}{cl} {1\over a(a+1)} (x^{a+1}- x- a(x-1)) & \mbox{~if~} a\ne 0, a\ne -1 \\ x\log x -x+1 & \mbox{~if~} a=0 \\ -\log x +x-1 & \mbox{~if~} a=-1\end{array}\right.
\end{equation}

Based on \eqref{eq9}, we get the minimum power-divergence estimator by

\begin{equation}\label{eq10} \hat{\bm \theta }_a := arg \min_{(\bm \lambda ,\bm \eta )\in \bm \Theta } D_{a}(\hat{\bf p}, {\bf p}(\bm \lambda ,\bm \eta )),
\end{equation}
where by $D_{a}(\hat{\bf p}, {\bf p}(\bm \lambda ,\bm \eta ))$ we denote $D_{\phi_a}(\hat{\bf p}, {\bf p}(\bm \lambda ,\bm \eta )),$ whose expression is

\begin{equation}\label{eq11} D_{a}(\hat{\bf p}, {\bf p}(\bm \lambda ,\bm \eta ))=\left\{ \begin{array}{cl} {1\over a(a+1)} {\displaystyle \sum_{j=1}^{2^k}}\left( {\hat{p}_j^{a+1}\over p_j(\bm \lambda ,\bm \eta )^a} -1 \right) & \mbox{~if~} a\ne 0, a\ne -1 \\  D_{Kullback}(\hat{\bf p}, {\bf p}(\bm \lambda ,\bm \eta )) & \mbox{~if~} a=0 \\ D_{Kullback}({\bf p}(\bm \lambda ,\bm \eta ), \hat{\bf p}) & \mbox{~if~} a=-1\end{array}\right.
\end{equation}

Observe that for $a=0$ we recover the MLE. If we look for the solution for different values of $a$, we obtain the results appearing in Table 1.

\begin{table}[h]
{\small
\begin{center}
\begin{tabular}{|c|ccccccccc|}
\hline Parameter / a & -1& -1/2 & 0 & 2/3 & 1 & 3/2 & 2 & 5/2 & 3 \\
\hline $\hat{\lambda}_1 $ & -2.3439 & -2.3436 & -2.3433 & -2.3429 & -2.3427 & -2.3424 & -2.3421 & -2.3418 & -2.3414 \\
$\hat{\lambda}_2 $ & 1.7194 & 1.7206 & 1.7219 & 1.7239 & 1.7251 & 1.7270 & 1.7291 & 1.7316 & 1.7343 \\
$\hat{\lambda}_3 $ & -0.8406 & -0.8405 & -0.8405 & -0.8404 & -0.8404 & -0.8403 & -0.8403 & -0.8403 & -0.8402 \\
$\hat{\lambda}_4 $ & 1.5710 & 1.5692 & 1.5675 & 1.5652 & 1.5642 & 1.5626 & 1.5611 & 1.5598 & 1.5585 \\
$\hat{\lambda}_5 $ & -2.0796 & -2.0753 & -2.0709 & -2.0648 & -2.0616 & -2.0567 & -2.0516 & -2.0462 & -2.0407 \\
$\hat{\lambda}_6 $ & 2.2989 & 2.2990 & 2.2991 & 2.2993 & 2.2994 & 2.2995 & 2.2997 & 2.2998 & 2.3000 \\
$\hat{\lambda}_7 $ & -0.9139 & -0.9132 & -0.9124 & -0.9114 & -0.9108 & -0.9100 & -0.9091 & -0.9081 & -0.9071 \\
$\hat{\lambda}_8 $ & 2.0116 & 2.0118 & 2.0121 & 2.0125 & 2.0128 & 2.0131 & 2.0135 & 2.0140 & 2.0144 \\
$\hat{\eta}_1 $ & 0.5026 & 0.5029 & 0.5041 & 0.5048 & 0.5060 & 0.5066 & 0.5067 & 0.5088 & 0.5095 \\
$\hat{\eta}_1 $ & 0.1674 & 0.1677 & 0.1689 & 0.1696	& 0.1708 & 0.1714 & 0.1713 & 0.1733 & 0.1737 \\
$\hat{\eta}_3 $ & -0.8722 & -0.8729 & -0.8728 & -0.8736	& -0.8731 & -0.8737 & -0.8749 & -0.8741 & -0.8748 \\
$\hat{\eta}_4 $ & -0.0040 & -0.0044 & -0.0039 & -0.0042 & -0.0036 & -0.0040 & -0.0050 & -0.0040 & -0.0047 \\
$\hat{p}_{(1,1)}$ & 0.0876 & 0.0876 & 0.0876 & 0.0876 & 0.0876 & 0.0877 & 0.0877 & 0.0877 & 0.0878 \\
$\hat{p}_{(1,2)}$ & 0.3014 & 0.3014 & 0.3014 & 0.3014 & 0.3015 & 0.3015 & 0.3015 & 0.3015 & 0.3015 \\
$\hat{p}_{(1,3)}$ & 0.1111 & 0.1115 & 0.1120 & 0.1126 & 0.1129 & 0.1134 & 0.1139 & 0.1144 & 0.1150 \\
$\hat{p}_{(1,4)}$ & 0.2862 & 0.2863 & 0.2865 & 0.2867 & 0.2868 & 0.2870 & 0.2872 & 0.2874 & 0.2876 \\
$\hat{p}_{(2,1)}$ & 0.0876 & 0.0876 & 0.0876 & 0.0876 & 0.0876 & 0.0877 & 0.0877 & 0.0877 & 0.0878 \\
$\hat{p}_{(2,2)}$ & 0.8279 & 0.8277 & 0.8274 & 0.8271 & 0.82670 & 0.8267 & 0.8265 & 0.8263 & 0.8261 \\
$\hat{p}_{(2,3)}$ & 0.1111 & 0.1115 & 0.1120 & 0.1126 & 0.1129 & 0.1134 & 0.1139 & 0.1144 & 0.1150 \\
$\hat{p}_{(2,4)}$ & 0.8820 & 0.8820 & 0.8821 & 0.8821 & 0.8821 & 0.8822 & 0.8822 & 0.8823 & 0.8823 \\
$\hat{p}_{(3,1)}$ & 0.8481 & 0.8482 & 0.8484 & 0.8486 & 0.8488 & 0.8490 & 0.8493 & 0.8496 & 0.8500 \\
$\hat{p}_{(3,2)}$ & 0.3014 & 0.3014 & 0.3014 & 0.3014 & 0.3015 & 0.3015 & 0.3015 & 0.3015 & 0.3015 \\
$\hat{p}_{(3,3)}$ & 0.9088 & 0.9088 & 0.9088 & 0.9088 & 0.9088 & 0.9088 & 0.9089 & 0.9089 & 0.9089 \\
$\hat{p}_{(3,4)}$ & 0.2862 & 0.2863 & 0.2865 & 0.2867 & 0.2868 & 0.2870 & 0.2872 & 0.2874 & 0.2876 \\
$\hat{p}_{(4,1)}$ & 0.8481 & 0.8482 & 0.8484 & 0.8486 & 0.8488 & 0.8490 & 0.8493 & 0.8496 & 0.8500 \\
$\hat{p}_{(4,2)}$ & 0.8279 & 0.8277 & 0.8274 & 0.8271 & 0.8270 & 0.8267 & 0.8265 & 0.8263 & 0.8261 \\
$\hat{p}_{(4,3)}$ & 0.9088 & 0.9088 & 0.9088 & 0.9088 & 0.9088 & 0.9088 & 0.9089 & 0.9089 & 0.9089 \\
$\hat{p}_{(4,4)}$ & 0.8820 & 0.8820 & 0.8821 & 0.8821 & 0.8821 & 0.8822 & 0.8822 & 0.8823 & 0.8823 \\
$\hat{w}_1 $ & 0.3890 & 0.3891 & 0.3892 & 0.3894 & 0.3895 & 0.3896 & 0.3898 & 0.3899 & 0.3901 \\
$\hat{w}_2 $ & 0.2782 & 0.2783 & 0.2784 & 0.2785 & 0.2785 & 0.2786 & 0.2787 & 0.2788 & 0.2789 \\
$\hat{w}_3 $ & 0.0984 & 0.0983 & 0.0982 & 0.0981 & 0.0981 & 0.0980 & 0.0979 & 0.0978 & 0.0977 \\
$\hat{w}_4 $ & 0.2344 & 0.2343 & 0.2342 & 0.2340 & 0.2339 & 0.2338 & 0.2337 & 0.2335 & 0.2333 \\
\hline
\end{tabular}
\end{center}
}
\caption{Power-divergence estimations for different values of $a$.}
\end{table}

It can be observed that there is not a significative difference among the obtained estimations for the different values of $a$; this agrees with the results developed in Section 3, where it is shown that the results coincide for big sample sizes, and in this sample we have $N=6658$.


\section{{\bf Simulation study}}

In this section we will carry out a simulation study to compare some alternative M$\phi $E with the MLE.

%

In order to do our simulation study we have considered a theoretical model with 5 dichotomous questions and 10 latent classes; next, 7 parameters $\lambda_j$ and 6 parameters $\eta_k$ are considered; the corresponding matrices of the model are

{\small
$$ {\bf Q}_1= \left( \begin{array}{ccccc}
                         1 & 0 & 0 & 0 & 0 \\
                         0 & 0 & 0 & 0 & 0 \\
                         0 & 0 & 0 & 0 & 0 \\
                         0 & 0 & 0 & 0 & 1 \\
                         0 & 0 & 0 & 1 & 0 \\
                         0 & 0 & 1 & 0 & 0 \\
                         0 & 1 & 0 & 0 & 0 \\
                         1 & 0 & 0 & 0 & 0 \\
                         0 & 0 & 0 & 1 & 0 \\
                         0 & 0 & 0 & 0 & 1 \\
\end{array}\right) ,
{\bf Q}_2 =  \left( \begin{array}{ccccc}
                         0 & 1 & 0 & 0 & 0 \\
                         1 & 0 & 0 & 0 & 0 \\
                         0 & 0 & 0 & 0 & 0 \\
                         0 & 0 & 0 & 0 & 0 \\
                         0 & 0 & 0 & 0 & 1 \\
                         0 & 0 & 0 & 1 & 0 \\
                         0 & 0 & 1 & 0 & 0 \\
                         0 & 0 & 0 & 0 & 1 \\
                         1 & 0 & 0 & 0 & 0 \\
                         0 & 0 & 1 & 0 & 0 \\
\end{array}\right) , {\bf Q}_3= \left( \begin{array}{ccccc}
                         0 & 0 & 1 & 0 & 0 \\
                         0 & 1 & 0 & 0 & 0 \\
                         1 & 0 & 0 & 0 & 0 \\
                         0 & 0 & 0 & 0 & 0 \\
                         0 & 0 & 0 & 0 & 0 \\
                         0 & 0 & 0 & 0 & 1 \\
                         0 & 0 & 0 & 1 & 0 \\
                         0 & 1 & 0 & 0 & 0 \\
                         0 & 0 & 0 & 0 & 1 \\
                         1 & 0 & 0 & 0 & 0 \\
\end{array}\right) , {\bf Q}_4=\left( \begin{array}{ccccc}
                         0 & 0 & 0 & 1 & 0 \\
                         0 & 0 & 1 & 0 & 0 \\
                         0 & 1 & 0 & 0 & 0 \\
                         1 & 0 & 0 & 0 & 0 \\
                         0 & 0 & 0 & 0 & 0 \\
                         0 & 0 & 0 & 0 & 0 \\
                         0 & 0 & 0 & 0 & 1 \\
                         0 & 0 & 0 & 0 & 0 \\
                         0 & 1 & 0 & 0 & 0 \\
                         0 & 0 & 0 & 0 & 0 \\
\end{array}\right) $$ }

$$ {\bf Q}_5=\left( \begin{array}{ccccc}
                         0 & 0 & 0 & 0 & 1 \\
                         0 & 0 & 0 & 1 & 0 \\
                         0 & 0 & 1 & 0 & 0 \\
                         0 & 1 & 0 & 0 & 0 \\
                         1 & 0 & 0 & 0 & 0 \\
                         0 & 0 & 0 & 0 & 0 \\
                         0 & 0 & 0 & 0 & 0 \\
                         0 & 0 & 1 & 0 & 0 \\
                         0 & 0 & 0 & 0 & 0 \\
                         0 & 0 & 0 & 1 & 0 \\
\end{array}\right) ,\, \, {\bf Q}_6= \left( \begin{array}{ccccc}
                         0 & 0 & 0 & 0 & 0 \\
                         0 & 0 & 0 & 0 & 1 \\
                         0 & 0 & 0 & 1 & 0 \\
                         0 & 0 & 1 & 0 & 0 \\
                         0 & 1 & 0 & 0 & 0 \\
                         1 & 0 & 0 & 0 & 0 \\
                         0 & 0 & 0 & 0 & 0 \\
                         0 & 0 & 0 & 0 & 0 \\
                         0 & 0 & 1 & 0 & 0 \\
                         0 & 1 & 0 & 0 & 0 \\
\end{array}\right) ,\, \, {\bf Q}_7=\left( \begin{array}{ccccc}
                         0 & 0 & 0 & 0 & 0 \\
                         0 & 0 & 0 & 0 & 0 \\
                         0 & 0 & 0 & 0 & 1 \\
                         0 & 0 & 0 & 1 & 0 \\
                         0 & 0 & 1 & 0 & 0 \\
                         0 & 1 & 0 & 0 & 0 \\
                         1 & 0 & 0 & 0 & 0 \\
                         0 & 0 & 0 & 1 & 0 \\
                         0 & 0 & 0 & 0 & 0 \\
                         0 & 0 & 0 & 0 & 0 \\
\end{array}\right) .$$ Matrix ${\bf C}$ is the null matrix. Matrix ${\bf V}$ is given by $$ {\bf V}= \left( \begin{array}{cccccc}
                         1 & 0 & 0 & 0 & 0 & 1 \\
                         1 & 0 & 0 & 0 & 0 & 0 \\
                         0 & 1 & 0 & 0 & 0 & 1 \\
                         0 & 1 & 0 & 0 & 0 & 0 \\
                         0 & 0 & 1 & 0 & 0 & 1 \\
                         0 & 0 & 1 & 0 & 0 & 0 \\
                         0 & 0 & 0 & 1 & 0 & 1 \\
                         0 & 0 & 0 & 1 & 0 & 0 \\
                         0 & 0 & 0 & 0 & 1 & 1 \\
                         0 & 0 & 0 & 0 & 1 & 0 \\
\end{array}\right) ,$$ while ${\bf d}= {\mathbf 0}.$ The theoretical values for vector $\bm \lambda $ and $\bm \eta $ are

$$ \bm \lambda_0 =(\lambda_1^0, ..., \lambda_7^0)=(-3, -2, -1, 0, 1, 2, 3),\, \, \bm \eta_0 =(\eta_1^0, ..., \eta_6^0)=(0.5, 1, 1.5, 2, 2.5, 3).$$

We shall consider the minimum power-divergence estimator defined in \eqref{eq10} with $D_{a}(\hat{\bf p}, {\bf p}(\bm \lambda ,\bm \eta ))$ defined in \eqref{eq11} with several different values $a$ ($a=-1, -{1\over 2}, 0, {2\over 3}, 1, {3\over 2}, 2, {5\over 2}, 3$) and several values of the sample size $N$ ($N=100, 200, 500, 1000, 2000$). For each combination $(N, a)$ we have conducted $n=1000$ random samples (simulations) in order to analyze the convergence of the algorithm. We want to estimate the $\bm \lambda , \bm \eta $ parameters and so ${\bf p}, {\bf w},$ by minimizing the $D_{\phi }(\hat{\bf p}, {\bf p}(\bm \lambda , \bm \eta ))$ function.

A previous analysis showed the existence of several points $\bm \lambda , \bm \eta $ with $\nabla D_{\phi }(\hat{\bf p}, {\bf p}(\bm \lambda , \bm \eta ))\approx 0$  and different values of $D_{\phi }(\hat{\bf p}, {\bf p}(\bm \lambda , \bm \eta ))$  function. In order to obtain a better approximation to a global minimum instead a local minimum or only a stationary point we apply the following multistart optimization algorithm: In step 1 we generate $N_{in}$ initial points (we used $N_{in}=500$); in step 2 we improve each point in their neighborhood using a low computational cost procedure;  if the improvement is satisfactory we proceed at the step 3 applying a good optimization algorithm from this improved point to obtain a local optimum and next we try to solve $\nabla D_{\phi }(\hat{\bf p}, {\bf p}(\bm \lambda , \bm \eta ))=0$ from this local optimum maintaining the decrease of $D_{\phi }(\hat{\bf p}, {\bf p}(\bm \lambda , \bm \eta ))$ function. The details of the algorithm are given next.

{\bf Step 1. Initialization.} Let

$$ R=\{ (\bm \lambda , \bm \eta ): \bm \lambda_{lo} \leq \bm \lambda \leq \bm \lambda^{up},\, \bm \eta_{lo} \leq \bm \eta \leq \bm \eta^{up}\} $$ be the region considered to seek the estimators. Let us denote by $N_{in}$ the number of points to be randomly generated in $R$. Initialize $D_{\phi }^{min}= +\infty , D_{\phi }^{in}= + \infty , i=1.$

{\bf Step 2. Rough improvement.} Generate $(\bm \lambda , \bm \eta )_i$ and perform a full iteration of a variant\footnote{The variant we used in step 2 consists in permuting randomly the $t+u$ parameters $(\bm \lambda , \bm \eta )$  for each initial point $i.$ The additional improvement consists in seeking a better point in the vector from the initial point to the final point obtained through the full iteration of the Hooke and Jeeves algorithm in double or half spacing steps towards exterior or interior relative to this vector. At most we need $2(t+u)+4$ evaluations of $D_{\phi }$. The criterion $D_{\phi}^{in}$ is used in order to discard non promising initial points from a finer and most costly improvement.} of the Hooke and Jeeves algorithm (Hooke $\& $ Jeeves (1961)) with an additional improvement step. Let $(\bar{\bm \lambda }, \bar{\bm \eta })_i$ be the point obtained through this procedure.

If $D_{\phi }(\hat{\bf p},(\bar{\bm \lambda }, \bar{\bm \eta })_i) < D_{\phi }^{in},$ set $D_{\phi }^{in} =D_{\phi }(\hat{\bf p},(\bar{\bm \lambda }, \bar{\bm \eta })_i)$  and go to step 3; otherwise, go to step 4.

{\bf Step 3. Fine improvement.} From $(\bar{\bm \lambda }, \bar{\bm \eta })_i$ as initial point, perform the limited memory quasi-Newton conjugate gradient algorithm (Gill $\& $ Murray (1979)) and let us denote by $(\bm \lambda ' , \bm \eta ')_i$   the point obtained through this procedure.

From $(\bm \lambda ', \bm \eta ')_i$ as initial point, solve the system $\nabla D_{\phi }(\hat{\bf p},(\bm \lambda , \bm \eta ))=0$ using the hybrid algorithm of Powell (Powell (1970)) and let us denote by $(\bm {\hat \lambda }, \bm {\hat \eta })_i$ the solution obtained through this procedure. If $D_{\phi }(\hat{\bf p},(\bm {\hat \lambda }, \bm {\hat \eta })_i)< D_{\phi }^{min},$  set $D_{\phi }^{min} = D_{\phi }(\hat{\bf p},(\bm {\hat \lambda }, \bm {\hat \eta })_i), \, \, (\bm \lambda , \bm \eta )^{min}=(\bm {\hat \lambda }, \bm {\hat \eta })_i $. Go to step 4.

{\bf Step 4. Stop.} If $i=N_{in},$ stop; otherwise let $i=i+1$ and go to step 2.

Part of the computations of this work were performed in EOLO, the HPC of Climate Change of the International Campus of Excellence (CEI) of Moncloa, funded by MECD and MICINN. This is a contribution of CEI Moncloa.

For simulation $l$ we get the values

$$ \hat{\lambda }^j_{a, l},\, j=1, ..., t,\, \, \, \, \hat{\eta }^k_{a, l},\, k=1, ..., u,$$ i.e. we obtain two vectors $\hat{\bm \lambda }^{(l)}_a =(\hat{\lambda }^1_{a, l}, ..., \hat{\lambda }^t_{a, l})$ and $\hat{\bm \eta }^{(l)}_a =(\hat{\eta }^1_{a, l}, ..., \hat{\eta }^u_{a, l})$ being $\hat{\bm \theta }^{(l)}_a=(\hat{\bm \lambda }^{(l)}_a, \hat{\bm \eta }^{(l)}_a)$ the minimum power-divergence estimator obtained for the $l$-th simulation using  $ D_{a}(\hat{\bf p}, {\bf p}(\bm \lambda ,\bm \eta )).$ Next, $\hat{\bm \theta }_a=(\hat{\bm \lambda }_a, \hat{\bm \eta }_a)$ is defined as

$$ \hat{\lambda }^j_a = {1\over n}\sum_{l=1}^{n} \hat{\lambda }^j_{a, l},\,  \, \,  \, \hat{\eta }^k_a = {1\over n}\sum_{l=1}^{n} \hat{\eta }^k_{a, l}.$$

For each $a$ we compute the mean squared error for each $\lambda_j$ and $\eta_k$

$$ mse(\lambda_j)={1\over n}\sum_{l=1}^{n}(\hat{\lambda }^j_a - \lambda_j^0)^2,\, \, \,  \, mse(\eta_k)={1\over n}\sum_{l=1}^{n}(\hat{\eta }^k_a - \eta_k^0)^2$$ and also the mean squared error for the random vectors $\hat{\bm \lambda }$ and $\hat{\bm \eta }_a$

$$ mse_{\bm \lambda }={1\over t} \sum_{j=1}^{t}mse(\lambda_j),\, \, \, \, mse_{\bm \eta }={1\over u} \sum_{k=1}^{u}mse(\eta_k).$$

In the different tables we present the values of $mse_{\bm \lambda }, mse_{\bm \eta }$ and also

$$ mse_{\bm \lambda , \bm \eta } = {1\over t+u}(t\, \,  mse(\bm \lambda ) + u\, \,  mse (\bm \eta ))$$ for each combination $(N, a).$ Similarly, we present the values of $mse_{\bf p}, mse_{\bf w}$ and

$$ mse_{\bf p, w} = {1\over j(i+1)}(ij \, \,  mse({\bf p}) + j\, \,  mse (\bf w)).$$

\begin{table}[h]
\begin{center}
\begin{tabular}{c|c|cccccc}
$N $      & $a$         & $ mse_{\bm \lambda }$ & $ mse_{\bm \eta }$ & $mse_{\bf p}$ & $mse_{\bf w}$ & $mse_{\bm \lambda , \bm \eta}$ & $mse_{\bf p, w}$ \\
\hline 100& -1  & 68.2008043 & 51.6399360 & 2.6187088 & 0.2414836 & 60.5573266 & 2.2225046 \\
          & -1/2& 31.4102753 & 21.7026774 & 0.5545327 & 0.0295127 & 26.9298455 & 0.4670294 \\
          & 0   & 27.6096591 & 21.5646935 & 0.5010663 & 0.0300314 & 24.8196749 & 0.4225605 \\
          & 2/3 & 26.1767552 & 21.5945653 & 0.4954186 & 0.0313723 & 24.0618983 & {\bf 0.4180775} \\
          & 1   & 25.3839485 & 21.7513194 & 0.4980696 & 0.0313950 & 23.7073504 & 0.4202905 \\
          & 3/2 & 24.7708020 & 21.4745848 & 0.5068482 & 0.0328157 & 23.2494710 & 0.4278428 \\
          & 2   & 24.3039758 & 22.1582747 & 0.5165476 & 0.0331497 & 23.3136522 & 0.4359813 \\
          & 5/2 & 24.4850663 & 21.7657641 & 0.5258098 & 0.0338732 & 23.2300038 & 0.4438204 \\
          & 3   & 23.9344458 & 21.3120512 & 0.5440716 & 0.0338817 & 22.7241099 & 0.4590400 \\

\hline 200& -1  & 52.3879545 & 41.3479516 & 1.5718160 & 0.1287486 & 47.2925685 & 1.3313048 \\
          & -1/2& 22.8239134 & 12.5410070 & 0.3112212 & 0.0129212 & 18.0779566 & 0.2615045 \\
          & 0   & 18.9598261 & 13.2682075 & 0.2714260 & 0.0127045 & 16.3329252 & 0.2283057 \\
          & 2/3 & 17.9741998 & 13.9441051 & 0.2640653 & 0.0132129 & 16.1141561 & {\bf 0.2222566} \\
          & 1   & 17.5365726 & 14.2143777 & 0.2663158 & 0.0134505 & 16.0032519 & 0.2241715 \\
          & 3/2 & 17.5724815 & 13.3321372 & 0.2710542 & 0.0137304 & 15.6153995 & 0.2281669 \\
          & 2   & 17.5617845 & 13.1178080 & 0.2760046 & 0.0142635 & 15.5107184 & 0.2323811 \\
          & 5/2 & 17.5915662 & 14.2253050 & 0.2830610 & 0.0145654 & 16.0379072 & 0.2383117 \\
          & 3   & 17.8876761 & 14.1583876 & 0.2901346 & 0.0149333 & 16.1664660 & 0.2442677 \\

\hline 500& -1  & 19.5850398 & 10.4099019 & 0.3351012 & 0.0180485 & 15.3503608 & 0.2822591 \\
          & -1/2& 11.8669496 & 4.5501940 & 0.1318223 & 0.0053937 & 8.4899855 & 0.1107509 \\
          & 0   & 10.8062705 & 4.0700143 & 0.1229104 & 0.0053519 & 7.6972292 & 0.1033173 \\
          & 2/3 & 10.7061006 & 4.4749985 & 0.1213370 & 0.0054439 & 7.8302073 & {\bf 0.1020215} \\
          & 1   & 10.7824896 & 4.3910689 & 0.1219875 & 0.0055265 & 7.8326031 & 0.1025774 \\
          & 3/2 & 10.9169580 & 4.9259907 & 0.1241744 & 0.0056426 & 8.1518962 & 0.1044191 \\
          & 2   & 11.0798328 & 5.0102359 & 0.1271099 & 0.0058042 & 8.2784804 & 0.1068922 \\
          & 5/2 & 11.3188481 & 4.6592600 & 0.1297526 & 0.0059495 & 8.2451920 & 0.1091187 \\
          & 3   & 11.5074602 & 5.1926346 & 0.1334828 & 0.0061370 & 8.5929253 & 0.1122585 \\

\hline 1000& -1 & 6.8372937 & 1.4649503 & 0.0779502 & 0.0033204 & 4.3577506 & 0.0655119 \\
           &-1/2& 6.1442723 & 0.9336013 & 0.0620760 & 0.0026275 & 3.7393472 & 0.0521679 \\
           & 0  & 6.2190043 & 1.1293547 & 0.0600988 & 0.0026132 & 3.8699353 & 0.0505178 \\
           &2/3 & 6.5774778 & 1.2108917 & 0.0599487 & 0.0026343 & 4.1005919 & {\bf 0.0503963} \\
           &1   & 6.7471263 & 0.9324688 & 0.0602965 & 0.0026541 & 4.0634382 & 0.0506894 \\
           &3/2 & 6.9406141 & 1.0615927 & 0.0611184 & 0.0026921 & 4.2272196 & 0.0513807 \\
           &2   & 7.1929378 & 1.0433170 & 0.0622090 & 0.0027387 & 4.3546513 & 0.0522973 \\
           &5/2 & 7.3457935 & 1.0575038 & 0.0634602 & 0.0027878 & 4.4435060 & 0.0533482 \\
           &3   & 7.5406717 & 1.1772223 & 0.0648446 & 0.0028436 & 4.6036950 & 0.0545111 \\

\hline 2000& -1 & 3.7138546 & 0.4670479 & 0.0316268 & 0.0012726 & 2.2153284 & 0.0265678 \\
           &-1/2& 3.8507852 & 0.3809852 & 0.0309693 & 0.0012662 & 2.2493391 & 0.0260188 \\
           & 0  & 3.9334490 & 0.3369390 & 0.0306378 & 0.0012655 & 2.2735213 & 0.0257424 \\
           &2/3 & 4.1309858 & 0.6180423 & 0.0305686 & 0.0012711 & 2.5096272 & {\bf 0.0256857} \\
           & 1  & 4.1953206 & 0.4477610 & 0.0306526 & 0.0012762 & 2.4656777 & 0.0257566 \\
           & 3/2& 4.3601170 & 0.3481949 & 0.0309031 & 0.0012868 & 2.5084606 & 0.0259670 \\
           & 2  & 4.4392610 & 0.4968572 & 0.0312681 & 0.0013003 & 2.6196900 & 0.0262735 \\
           & 5/2& 4.5773326 & 0.3998131 & 0.0317068 & 0.0013157 & 2.6492467 & 0.0266417 \\
           & 3  & 4.6916695 & 0.6705992 & 0.0322143 & 0.0013330 & 2.8357909 & 0.0270675 \\
\hline
\end{tabular}
\end{center}
\caption{mse of the simulation study $n=1000$}
\end{table}

\begin{table}[h]
\begin{center}
\begin{tabular}{c|c|cccccc}
$N $      & $a$         & $ bias_{\bm \lambda }$ & $ bias_{\bm \eta }$ & $bias_{\bf p}$ & $bias_{\bf w}$ & $bias_{\bm \lambda , \bm \eta}$ & $bias_{\bf p, w}$ \\
\hline 100& -1  & 39.2071730 & 26.0542812 & 0.4699344 & 0.0336229 & 33.1366075 & 0.3972158 \\
          &-1/2 & 9.0358383 & 4.7778114 & 0.0343787 & 0.0003508 & 7.0705951 & 0.0287074 \\
          &0    & 5.4738807 & 3.8825077 & 0.0061595 & 0.0003859 & 4.7394009 & {\bf 0.0051972} \\
          &2/3  & 4.3687819 & 3.6680702 & 0.0091762 & 0.0009732 & 4.0453765 & 0.0078091 \\
          & 1   & 3.9599698 & 3.6716953 & 0.0127112 & 0.0011208 & 3.8269200 & 0.0107795 \\
          &3/2  & 3.6450364 & 3.7246612 & 0.0192119 & 0.0016214 & 3.6817863 & 0.0162802 \\
          &2    & 3.4295955 & 3.5320425 & 0.0241503 & 0.0018552 & 3.4768788 & 0.0204344 \\
          &5/2  & 3.3988379 & 3.5071508 & 0.0281768 & 0.0020897 & 3.4488285 & 0.0238289 \\
          & 3   & 3.1381705 & 3.3378409 & 0.0329561 & 0.0021393 & 3.2303261 & 0.0278200 \\

\hline 200& -1  & 25.7946436 & 14.5671733 & 0.1497812 & 0.0084382 & 20.6127342 & 0.1262241 \\
          & -1/2& 5.5092624 & 1.8020270 & 0.0197433 & 0.0001067 & 3.7982307 & 0.0164705 \\
          & 0   & 3.4967022 & 1.8492188 & 0.0050793 & 0.0001259 & 2.7363252 & 0.0042537 \\
          &2/3  & 3.1998206 & 1.8295172 & 0.0023832 & 0.0002782 & 2.5673729 & {\bf 0.0020324} \\
          &1    & 3.0586797 & 1.8898579 & 0.0030349 & 0.0003635 & 2.5192235 & 0.0025896 \\
          &3/2  & 3.0801905 & 1.7332743 & 0.0048118 & 0.0004821 & 2.4585369 & 0.0040902 \\
          &2    & 3.0828546 & 1.7395245 & 0.0067276 & 0.0006038 & 2.4628561 & 0.0057070 \\
          &5/2  & 3.0397273 & 1.9315050 & 0.0089581 & 0.0007046 & 2.5282401 & 0.0075825 \\
          & 3   & 3.0581915 & 1.8843475 & 0.0109454 & 0.0008041 & 2.5164173 & 0.0092552 \\

\hline 500& -1  & 4.3981249 & 0.9892542 & 0.0086254 & 0.0001784 & 2.8248000 & 0.0072176 \\
          &-1/2 & 1.7577971 & 0.2891266 & 0.0041876 & 0.0000585 & 1.0799492 & 0.0034995 \\
          & 0   & 1.5854943 & 0.2555561 & 0.0015625 & 0.0000760 & 0.9716767 & 0.0013147 \\
          &2/3  & 1.6593839 & 0.2641776 & 0.0010594 & 0.0001179 & 1.0154426 & {\bf 0.0009025} \\
          &1    & 1.7048907 & 0.2539989 & 0.0012978 & 0.0001439 & 1.0352484 & 0.0011055 \\
          &3/2  & 1.7740016 & 0.2701422 & 0.0019558 & 0.0001855 & 1.0799127 & 0.0016607 \\
          &2    & 1.8610774 & 0.3141347 & 0.0028182 & 0.0002319 & 1.1471038 & 0.0023872 \\
          &5/2  & 1.9439342 & 0.3046043 & 0.0037351 & 0.0002731 & 1.1873204 & 0.0031581 \\
          &3    & 2.0118778 & 0.3579726 & 0.0047165 & 0.0003225 & 1.2485369 & 0.0039841 \\

\hline 1000& -1 & 0.8113274 & 0.0356155 & 0.0024739 & 0.0000203 & 0.4533065 & 0.0020650 \\
           &-1/2& 0.7869947 & 0.0229321 & 0.0011349 & 0.0000105 & 0.4343504 & 0.0009475 \\
           &0   & 0.8827217 & 0.0221178 & 0.0004414 & 0.0000121 & 0.4855199 & 0.0003699 \\
           &2/3 & 1.0245992 & 0.0200585 & 0.0001346 & 0.0000184 & 0.5609650 & {\bf 0.0001153} \\
           & 1  & 1.0862811 & 0.0187584 & 0.0001361 & 0.0000229 & 0.5935783 & 0.0001172 \\
           & 3/2& 1.1806177 & 0.0175534 & 0.0002631 & 0.0000310 & 0.6438188 & 0.0002244 \\
           &2   & 1.2696392 & 0.0183709 & 0.0004966 & 0.0000406 & 0.6921308 & 0.0004206 \\
           &5/2 & 1.3439794 & 0.0211222 & 0.0008046 & 0.0000514 & 0.7334299 & 0.0006791 \\
           & 3  & 1.4235789 & 0.0205671 & 0.0011575 & 0.0000631 & 0.7760350 & 0.0009751 \\

\hline 2000& -1 & 0.3974134 & 0.0033490 & 0.0003486 & 0.0000019 & 0.2155375 & 0.0002908 \\
           &-1/2& 0.4352462 & 0.0032241 & 0.0001612 & 0.0000024 & 0.2358514 & 0.0001347 \\
           &0   & 0.4677781 & 0.0024132 & 0.0000535 & 0.0000037 & 0.2529943 & 0.0000452 \\
           &2/3 & 0.5280862 & 0.0019664 & 0.0000162 & 0.0000064 & 0.2852616 & {\bf 0.0000146} \\
           & 1  & 0.5536678 & 0.0018305 & 0.0000349 & 0.0000081 & 0.2989736 & 0.0000305 \\
           &3/2 & 0.6022356 & 0.0019032 & 0.0000999 & 0.0000111 & 0.3251591 & 0.0000851 \\
           & 2  & 0.6391803 & 0.0016763 & 0.0002003 & 0.0000146 & 0.3449476 & 0.0001693 \\
           &5/2 & 0.6836233 & 0.0030145 & 0.0003271 & 0.0000183 & 0.3694962 & 0.0002756 \\
           & 3  & 0.7225695 & 0.0028221 & 0.0004746 & 0.0000225 & 0.3903784 & 0.0003992 \\
\hline
\end{tabular}
\end{center}
\caption{Bias of the simulation study $n=1000$}
\end{table}

In Table 2 we have considered the mean quadratic error for the estimation of $\bm \lambda , \bm \eta , (\bm \lambda , \bm \eta ),$ as well as for the estimations of $\bf{p}, \bf{w}$ and $(\bf{p}, \bf{w})$. In Table 3, the bias corresponding to $\bm {\hat \lambda }, \bm {\hat \eta }, (\bm {\hat \lambda }, \bm {\hat \eta })$ as well as $\bf{\hat{p}}, \bf{\hat{w}}, (\bf{\hat{p}},\bf{\hat{w}})$ are presented. We cannot forget that in a latent model for binary data the interest focuses on $\bf{p}$ and $\bf{w}$. Therefore, we pay special attention to the jointly estimation of these parameters, namely $(\bf{\hat{p}},\bf{\hat{w}}).$ From these tables, we might infer some conclusions:

\begin{itemize}
\item In accordance with our theoretical results, when $N$ increases, both the mean squared errors and the bias decrease and the bias tends to zero. It is also interesting to observe that for the negative value of $a$ convergence is slower than for the positive values of $a$.
\item If we observe Table 2 we can see that for all sample sizes ($N$) considered, the value of $mse_{\bf w}$ is smaller for $a=0$ (MLE) than for $a=2/3;$ while $mse_{\bf p}$ is smaller for $a=2/3$. Thus, for a strict point of view the recommendation should be to use the MLE when the interest lays on ${\bf w}$ and take $a=2/3$ when the focus is on ${\bf p}$. On the other hand, in order to give a unified criterion of estimation we have considered $mse_{\bf p, w}$ as a global measure of the mean squared error for ${\bf p}$ and ${\bf w}$. Considering $mse_{\bf p, w}$, the values associated to $a=2/3$ are better than the values associated to $a=0$ (MLE). Then, it seems that the amount of the increase of $mse_{\bf w}$ for $a=2/3$ is smaller than the amount of the increase of $mse_{\bf p}$ for $a=0$. Based on the previous comments and taking into account the behavior of the mean squared error, we recommend $a=2/3.$ In Table 3 we present the results corresponding to the bias and the conclusions are similar to those of $mse_{\bf p, w}$, (except for $N=100$, in which $bias_{\bf p, w}$ associated to the MLE is smaller than $bias_{\bf p, w}$ for $a=2/3$) and we can derive the same conclusions as before.

    We can also observe that the M$\phi $E obtained for $a=1$ has in general a better behavior than the MLE estimator in terms of bias and in terms of mean square error. Thus, we have obtained two values that seem to work better than the MLE. Notice that $a=2/3$ corresponds to the value proposed by Cressie in the context of testing goodness-of-fit and $a=1$ corresponds to the minimum $\chi^2$ estimator considered by many authors in different statistical problems. At this point we refer to the paper of Berkson (1980).
\end{itemize}

The previous results pointed out the efficiency of the minimum power divergence for $a=2/3$ for latent class models for binary data; on the other hand, it is also important to study the infinitesimal robustness. To deal with this point, we have carried out a simulation study in which we compare the robustness of the MLE with the robustness of the minimum power divergence estimator for $a=2/3.$

We denote by $M$ the latent model with binary data considered in this section in our simulation study. Next, we define the latent models $M_j, j=1, ..., 6,$ in which we add a new parameter $\lambda_8$ and assign it different values (0.5, 1, 1.5, 2, -0.5, -1). Matrix ${\bf Q}_8$ is given by

$$ {\bf Q}_8= \left( \begin{array}{ccccc}
                         1 & 1 & 1 & 1 & 1 \\
                         1 & 1 & 1 & 1 & 1 \\
                         1 & 1 & 1 & 1 & 1 \\
                         1 & 1 & 1 & 1 & 1 \\
                         1 & 1 & 1 & 1 & 1 \\
                         0 & 0 & 0 & 0 & 0 \\
                         0 & 0 & 0 & 0 & 0 \\
                         0 & 0 & 0 & 0 & 0 \\
                         0 & 0 & 0 & 0 & 0 \\
                         0 & 0 & 0 & 0 & 0 \\
\end{array}\right) . $$

We consider the mixed contaminated latent model with binary data defined by

$$ L_j= (1-\epsilon )M + \epsilon M_j, j=1, ..., 6,$$ where $\epsilon $ represents the ratio of contamination. Applying the algorithm developed in this section, for these models we obtain both the maximum likelihood estimator and the minimum power divergence estimator for $a=2/3$ when the observations are independently drawn for models $L_j.$ We consider again the sample sizes $N=100, 200, 500, 1000$ and $2000.$ Finally, we take the ratio of contamination $\epsilon =0.05.$ In next table we present the mean squared error for jointly ${\bf p}$ and ${\bf w},$ i.e. $mse_{\bf p, w}.$

{\small
\begin{center}
\begin{tabular}{|c|c|cccccc|}
\hline $N$ & & $M_1= 0.5$ & $M_2=1$ & $M_3=1.5$ & $M_4=2$ & $M_5=-0.5$ & $M_6=-1.5$ \\
\hline 100 & \begin{tabular}{c} $a=0$ \\ $a=2/3$ \end{tabular} & \begin{tabular}{c} 0.4171233\\ 0.4067895 \end{tabular} & \begin{tabular}{c} 0.4231311\\ 0.4128136 \end{tabular} & \begin{tabular}{c} 0.4198078\\ 0.4054495 \end{tabular} & \begin{tabular}{c} 0.4242797\\ 0.4191912 \end{tabular} & \begin{tabular}{c} 0.4421665\\ 0.4362660 \end{tabular} & \begin{tabular}{c} 0.4424415\\ 0.4385473 \end{tabular} \\
\hline 200 & \begin{tabular}{c} $a=0$ \\ $a=2/3$ \end{tabular} & \begin{tabular}{c} 0.2435659 \\ 0.2364162 \end{tabular} & \begin{tabular}{c} 0.2428468 \\ 0.2352565 \end{tabular} & \begin{tabular}{c} 0.2381769 \\ 0.2275114 \end{tabular} & \begin{tabular}{c} 0.2423141 \\ 0.2368489 \end{tabular} & \begin{tabular}{c} 0.2520425 \\ 0.2445993 \end{tabular} & \begin{tabular}{c} 0.2528909 \\ 0.2435447 \end{tabular}\\
\hline 500 & \begin{tabular}{c} $a=0$ \\ $a=2/3$ \end{tabular} & \begin{tabular}{c} 0.1008650 \\ 0.0991166 \end{tabular} & \begin{tabular}{c} 0.1019014 \\ 0.1004413 \end{tabular} & \begin{tabular}{c} 0.1000709 \\ 0.0975282 \end{tabular} & \begin{tabular}{c} 0.1016192 \\ 0.0993123 \end{tabular} & \begin{tabular}{c} 0.1032649 \\ 0.1005406 \end{tabular} & \begin{tabular}{c} 0.1105525 \\ 0.1091972 \end{tabular}\\
\hline 1000 & \begin{tabular}{c} $a=0$ \\ $a=2/3$ \end{tabular} & \begin{tabular}{c} 0.0498078 \\ 0.0494964 \end{tabular} & \begin{tabular}{c} 0.0495336 \\ 0.0491904 \end{tabular} & \begin{tabular}{c} 0.0490689 \\ 0.0487624 \end{tabular} & \begin{tabular}{c} 0.0527398 \\ 0.0519612 \end{tabular} & \begin{tabular}{c} 0.0508746 \\ 0.0507482 \end{tabular} & \begin{tabular}{c} 0.0540802\\ 0.0537044 \end{tabular} \\
\hline 2000 & \begin{tabular}{c} $a=0$ \\ $a=2/3$ \end{tabular} & \begin{tabular}{c} 0.0254735 \\ 0.0254254 \end{tabular} & \begin{tabular}{c} 0.0249197 \\ 0.0248472 \end{tabular} & \begin{tabular}{c} 0.0270387 \\ 0.0269094 \end{tabular} & \begin{tabular}{c} 0.0266822 \\ 0.0266083 \end{tabular} & \begin{tabular}{c} 0.0249668 \\ 0.0249112 \end{tabular} & \begin{tabular}{c} 0.0278638 \\ 0.0278340 \end{tabular} \\
\hline
\end{tabular}
\end{center}
}

We can observe that for all the models under consideration the minimum power divergence estimator with $a=2/3$ has a better behavior than the MLE ($a=0$). Then, based on this simulation study we can say that the minimum power divergence estimator with $a=2/3$ works better than the MLE in the sense of the infinitesimal robustness.

\section{{\bf Conclusions}}

From a classical point of view the unknown parameters in LCM for binary data have been estimated using MLE. In this paper, using the parametrization for LCM with binary data proposed in Formann (1985), we introduce and study the family of M$\phi $E, paying especial attention to the family of power-divergence estimators. This family of estimators can be considered as an extension of the MLE in the sense that the MLE is an estimator included in this family. For any M$\phi $E, we have obtained its asymptotic distribution, showing that is the same for all of them and the same as MLE, i.e. it does not depend on the function $\phi $ under consideration. It is shown that these estimators are BAN (Best asymptotically normal estimators) and that they should behave in the same way for big sample sizes; moreover, their speed of convergence is the same. In this sense, we can observe in the numerical example of Section 4 that the estimations obtained for the different parameters are quite similar because the sample size ($N=6658$) is big enough to apply the asymptotical results.

The asymptotic results do not provide information about the behavior of the different estimators while dealing with sample sizes that are not big enough. Usually, the way to study the behavior for such sample sizes is through a simulation study. In general, it is not easy to establish when a sample size cannot be considered big enough to apply the asymptotic results, as this will depend on the number of parameters to be estimated. In our case, this is not a problem because for all the sample sizes under  consideration the estimations for $a=2/3$ have a better behavior than for $a=0$ (MLE). To study this point, we have carried out a simulation study; this simulation study seems to show that there are indeed differences when the sample size is not big enough to apply the asymptotical results for different estimators under consideration. From this study, we have seen that the minimum power divergence estimator for $a=2/3$ and the minimum $\chi^2$ estimator exhibit a better behavior than the MLE in LCM for binary data. We recommend the use of minimum power divergence estimator for $a=2/3$ for LCM for binary data under the parametrization given by Formann (1985) and presented in \eqref{F1} and \eqref{F2}. A small simulation study carried out that seems to mean that the minimum power divergence estimator obtained for $a=2/3$ has a better behavior, in the sense of robustness, than the MLE.

\section{{\bf Acknowledgements}}

We thank the anonymous referees for their comments and remarks that have improved the new version of the paper.
%

\section*{{\bf Appendix}}

{\bf Remark 1}

We are going to develop the calculations for ${\displaystyle{\partial p({\bf y_{\nu }}, \bm \lambda , \bm \eta )\over \partial \lambda_{\alpha }}}$ and ${\displaystyle{\partial p({\bf y_{\nu }}, \bm \lambda , \bm \eta )\over \partial \eta_{\beta }}}.$

For ${\displaystyle{\partial p({\bf y_{\nu }}, \bm \lambda , \bm \eta )\over \partial \lambda_{\alpha }}}$ note that

\begin{eqnarray*}
p({\bf y_{\nu }}, \bm \lambda , \bm \eta ) & = & \sum_{j=1}^m w_j \prod_{i=1}^k \left( {exp ({\displaystyle \sum_{r=1}^t q_{jir} \lambda_r + c_{ji}})\over 1 +exp ({\displaystyle \sum_{r=1}^t q_{jir} \lambda_r + c_{ji}})}\right)^{y_{\nu i}} \left( 1- {exp ({\displaystyle \sum_{r=1}^t q_{jir} \lambda_r + c_{ji}})\over 1 +exp ({\displaystyle \sum_{r=1}^t q_{jir} \lambda_r + c_{ji}})} \right)^{1-y_{\nu i}} \\
& = & \sum_{j=1}^m w_j \prod_{i=1}^k {exp \left( y_{\nu i}\left( {\displaystyle \sum_{r=1}^t q_{jir} \lambda_r + c_{ji}}\right) \right) \over 1 +exp ({\displaystyle \sum_{r=1}^t q_{jir} \lambda_r + c_{ji}}) }
\end{eqnarray*}

Now,

\begin{eqnarray*}
{\partial \left( {exp \left( y_{\nu i}\left( {\displaystyle \sum_{r=1}^t q_{jir} \lambda_r + c_{ji}}\right) \right) \over 1 +exp ({\displaystyle \sum_{r=1}^t q_{jir} \lambda_r + c_{ji}})} \right) \over \partial \lambda_{\alpha }} & = & {exp \left( y_{\nu i}\left( {\displaystyle \sum_{r=1}^t q_{jir} \lambda_r + c_{ji}}\right) \right)  q_{ji\alpha }\over 1 +exp ({\displaystyle \sum_{r=1}^t q_{jir} \lambda_r + c_{ji}})} \left[ y_{\nu i} - {exp \left( {\displaystyle \sum_{r=1}^t q_{jir} \lambda_r + c_{ji}}\right) \over 1 +exp ({\displaystyle \sum_{r=1}^t q_{jir} \lambda_r + c_{ji}})}\right] \\
& = & {exp \left( y_{\nu i}\left( {\displaystyle \sum_{r=1}^t q_{jir} \lambda_r + c_{ji}}\right) \right) \over 1 +exp ({\displaystyle \sum_{r=1}^t q_{jir} \lambda_r + c_{ji}})} q_{ji\alpha }\left( y_{\nu i} - p_{ji}\right) ,
\end{eqnarray*}
whence

$$ {\partial p({\bf y_{\nu }}, \bm \lambda , \bm \eta )\over \partial \lambda_{\alpha }} = \sum_{j=1}^m w_j Pr ({\bf y_{\nu }} | P_{\nu }\in C_j ) \sum_{i=1}^k q_{ji\alpha } (y_{\nu i}- p_{ji}), \, \, \alpha =1, ..., t. $$

Similarly,

$$ p({\bf y_{\nu }}, \bm \lambda , \bm \eta ) = \sum_{j=1}^m {exp ({\displaystyle \sum_{r=1}^u v_{jr} \eta_r + d_{j}})\over {\displaystyle \sum_{j=1}^m exp (\sum_{r=1}^u v_{jr} \eta_r + d_{j})}} Pr ({\bf y_{\nu }} | P_{\nu }\in C_j ).$$

Now,

\begin{eqnarray*}
{\partial \left( {exp ({\displaystyle \sum_{r=1}^u v_{jr} \eta_r + d_{j}})\over {\displaystyle \sum_{h=1}^m exp (\sum_{r=1}^u v_{hr} \eta_r + d_{h})}} \right) \over \partial \eta_{\beta} } & = & {exp ({\displaystyle \sum_{r=1}^u v_{jr} \eta_r + d_{j}})\over {\displaystyle \sum_{h=1}^m exp (\sum_{r=1}^u v_{hr} \eta_r + d_{h})}} \left[ v_{j\beta } - {{\displaystyle \sum_{h=1}^m exp (\sum_{r=1}^u v_{hr} \eta_r + d_{h}})v_{h\beta }\over {\displaystyle \sum_{h=1}^m exp (\sum_{r=1}^u v_{hr} \eta_r + d_{h})}}\right] \\
& = & w_j \left[ v_{j\beta } - \sum_{h=1}^m w_h v_{h\beta }\right] ,
\end{eqnarray*}
whence
$$ {\partial p({\bf y_{\nu }}, \bm \lambda , \bm \eta )\over \partial \eta_{\beta }} = \sum_{j=1}^m w_j Pr ({\bf y_{\nu }} | P_{\nu }\in C_j )\left[ v_{j\beta } - \sum_{h=1}^m w_h v_{h\beta } \right] ,\, \, \beta =1, ..., u.$$

{\bf Proof of Theorem \ref{teo1}.}

Let $l^{2^k}$ be the interior of the $2^k$-dimensional unit cube; then, the interior of $\Delta_{2^k}$ is contained in $l^{2^k}.$ Let $W$ be a neighborhood of $(\bm \lambda_0, \bm \eta_0 ),$ the true value of the unknown parameter $(\bm \lambda , \bm \eta ),$ on which

\begin{eqnarray*}
{\bf p}: & \Theta & \rightarrow \Delta_{2^k} \\
    & (\bm \lambda , \bm \eta ) & \mapsto  {\bf p}(\bm \lambda , \bm \eta ):= (p_1(\bm \lambda , \bm \eta ), ..., p_{2^k} (\bm \lambda , \bm \eta ))
\end{eqnarray*}
has continuous second partial derivatives. Let

$$ {\bf F}:=(F_1, ..., F_{t+u}): l^{2^k} \times W \rightarrow \mathbb{R}^{t+u}$$ whose components $F_j,\, j=1, ..., t+u$ are defined by

$$ F_j(\tilde{p}_1, ..., \tilde{p}_{2^k}; \lambda_1, ..., \lambda_t; \eta_1, ..., \eta_u ):= {\partial D_{\phi }({\bf \tilde{p}}, {\bf p}(\bm \lambda , \bm \eta )) \over \partial s_j},\, j=1, ..., t+u,$$ where $s_j$ is defined in \eqref{eq8}.

It holds

$$ F_j(p_1(\bm \lambda_0 , \bm \eta_0 ), ..., p_{2^k}(\bm \lambda_0 , \bm \eta_0 );\lambda_1^0, ..., \lambda_t^0; \eta_1^0, ..., \eta_u^0)=0,\, \forall j=1, ..., t+u$$ due to

$$ {\partial D_{\phi } ({\bf \tilde{p}}, {\bf p}(\bm \lambda , \bm \eta ))\over \partial \lambda_{\alpha }}= \sum_{\nu =1}^{2^k} \left\{ \phi \left( {\tilde{p}_{\nu }\over p_{\nu }(\bm \lambda , \bm \eta )} \right) - {\tilde{p}_{\nu }\over p_{\nu }(\bm \lambda , \bm \eta )} \phi '\left( {\tilde{p}_{\nu }\over p_{\nu }(\bm \lambda , \bm \eta )} \right) \right\} {\partial p_{\nu }( \bm \lambda , \bm \eta )\over \partial \lambda_{\alpha }}  ,\, \alpha =1, ..., t.$$

$$ {\partial D_{\phi } ({\bf \hat{p}}, {\bf p}(\bm \lambda , \bm \eta ))\over \partial \eta_{\beta }}= \sum_{\nu =1}^{2^k} \left\{ \phi \left( {\tilde{p}_{\nu }\over p_{\nu }(\bm \lambda , \bm \eta )} \right) - {\tilde{p}_{\nu }\over p_{\nu }(\bm \lambda , \bm \eta )} \phi '\left( {\tilde{p}_{\nu }\over p_{\nu }(\bm \lambda , \bm \eta )} \right)  \right\} {\partial p_{\nu }(\bm \lambda , \bm \eta )\over \partial \eta_{\beta }} ,\, \beta =1, ..., u.$$

In the following we shall rewrite the two previous expressions by

$$ {\partial D_{\phi } ({\bf \tilde{p}}, {\bf p}(\bm \lambda , \bm \eta ))\over \partial s_{j}},\, j=1, ..., t+u.$$

Since

\begin{eqnarray*}
{\partial \over \partial s_r}\left( {\partial D_{\phi } ({\bf \tilde{p}}, {\bf p}(\bm \lambda , \bm \eta ))\over \partial s_{j}} \right) & = & -\sum_{\nu =1}^{2^k} \phi '\left( {\tilde{p}_{\nu }\over p_{\nu }(\bm \lambda , \bm \eta )} \right)  {\tilde{p}_{\nu }\over p_{\nu }(\bm \lambda , \bm \eta )^2}
{\partial p_{\nu }(\bm \lambda , \bm \eta )\over \partial s_{r}} {\partial p_{\nu }(\bm \lambda , \bm \eta )\over \partial s_{j}} \\ & &
+ \sum_{\nu =1}^{2^k} \phi ''\left( {\tilde{p}_{\nu }\over p_{\nu }(\bm \lambda , \bm \eta )} \right)  {\tilde{p}_{\nu }\over p_{\nu }(\bm \lambda , \bm \eta )^2}
{\partial p_{\nu }(\bm \lambda , \bm \eta )\over \partial s_{r}} {\partial p_{\nu }(\bm \lambda , \bm \eta )\over \partial s_{j}} {\tilde{p}_{\nu }\over p_{\nu }(\bm \lambda , \bm \eta )}\\ & & + \sum_{\nu =1}^{2^k} \phi '\left( {\tilde{p}_{\nu }\over p_{\nu }(\bm \lambda , \bm \eta )} \right)  {\tilde{p}_{\nu }\over p_{\nu }(\bm \lambda , \bm \eta )^2}
{\partial p_{\nu }(\bm \lambda , \bm \eta )\over \partial s_{r}} {\partial p_{\nu }(\bm \lambda , \bm \eta )\over \partial s_{j}} \\ & &  +
\sum_{\nu =1}^{2^k} {\partial^2 p_{\nu }(\bm \lambda , \bm \eta )\over \partial s_{r} s_j} \left\{ \phi \left( {\tilde{p}_{\nu }\over p_{\nu }(\bm \lambda , \bm \eta )} \right)  - \phi '\left( {\tilde{p}_{\nu }\over p_{\nu }(\bm \lambda , \bm \eta )} \right) {\tilde{p}_{\nu }\over p_{\nu }(\bm \lambda , \bm \eta )}\right\} ,
\end{eqnarray*}
and denoting $\pi_{\nu }=p_{\nu }(\bm \lambda_0 , \bm \eta_0), \nu =1, ..., 2^k,$ the $(t+u)\times (t+u)$ matrix ${\bf J}_{\bf F}$ associated with function ${\bf F}$ at point $({\bf p}(\bm \lambda_0 , \bm \eta_0), (\bm \lambda_0 , \bm \eta_0 ))$ is given by

\begin{eqnarray*}
{\partial {\bf F}\over \partial (\bm \lambda_0 , \bm \eta_0 )} & = & \left( {\partial {\bf F}\over \partial (\bm \lambda , \bm \eta )} \right)_{({\bf \tilde{p}}, (\bm \lambda , \bm \eta ))=(\pi_1, ..., \pi_{2^k}; \lambda_1^0, ..., \lambda_t^0; \eta_1^0, ..., \eta_u^0)} \\
& = & \left( \left( {\partial \over \partial s_r}\left( {\partial D_{\phi } ({\bf \tilde{p}}, {\bf p}(\bm \lambda , \bm \eta ))\over \partial s_{j}} \right) \right)_{\stackrel{j=1, ..., t+u}{r=1, ..., t+u}} \right)_{({\bf \tilde{p}}, (\bm \lambda , \bm \eta ))=(\pi_1, ..., \pi_{2^k}; \lambda_1^0, ..., \lambda_t^0; \eta_1^0, ..., \eta_u^0)} \\
& = & \phi ''(1) \left( \sum_{l=1}^{2^k} {1\over p_l(\bm \lambda_0 , \bm \eta_0 )}{\partial p_l(\bm \lambda_0 , \bm \eta_0 )\over \partial s_{r}} {\partial p_l(\bm \lambda_0 , \bm \eta_0 )\over \partial s_{j}} \right)_{\stackrel{j=1, ..., t+u}{r=1, ..., t+u}}
\end{eqnarray*}

To get the last expression we are using that $\phi (1)=\phi '(1)=0.$ Recall that if ${\bf B}$ is a $p\times q$ matrix with $rank({\bf B})=p$ and ${\bf C}$ is a $q\times s$ matrix with $rank({\bf C})=q,$ then $rank({\bf BC})=p.$ Taking

$$ {\bf B}= {\bf J}(\bm \lambda_0 , \bm \eta_0)^T, \, \, {\bf C}={\bf D}_{{\bf p}(\bm \lambda_0 , \bm \eta_0)}^{-{1\over 2}},$$ it follows that ${\bf A}(\bm \lambda_0 , \bm \eta_0)^T={\bf BC}$ has rank $t+u$ applying the fourth condition of Birch. Also,

$$ rank ({\bf A}(\bm \lambda_0 , \bm \eta_0)^T{\bf A}(\bm \lambda_0 , \bm \eta_0))=rank ({\bf A}(\bm \lambda_0 , \bm \eta_0){\bf A}(\bm \lambda_0 , \bm \eta_0)^T)=rank ({\bf A}(\bm \lambda_0 , \bm \eta_0))=t+u.$$

Therefore, the $(t+u)\times (t+u)$ matrix ${\partial {\bf F}\over \partial (\bm \lambda_0 , \bm \eta_0)}$ is nonsingular at $(\pi_1, ..., \pi_{2^k}; \lambda_1^0, ..., \lambda_t^0; \eta_1^0, ..., \eta_u^0)$.

Applying the Implicit Function Theorem, there exists a neighborhood $U$ of $({\bf p}(\bm \lambda_0 , \bm \eta_0), (\bm \lambda_0 , \bm \eta_0 ))$ such that the matrix ${\bf J}_{\bf F}$ is nonsingular (in our case ${\bf J}_{\bf F}$ at $({\bf p}(\bm \lambda_0 , \bm \eta_0), (\bm \lambda_0 , \bm \eta_0 ))$ is positive definite and then it is continuously differentiable). Also, there exists a continuously differentiable function

$$\tilde{\bm \theta }:A\subset l^{2^k} \rightarrow \mathbb{R}^{t+u} $$ such that ${\bf p}(\bm \lambda_0 , \bm \eta_0)\in A$ and

\begin{equation}\label{eq6} \left\{ ({\bf \tilde{p}}, (\bm \lambda , \bm \eta ))\in U : {\bf F}({\bf \tilde{p}}, (\bm \lambda , \bm \eta ))=0\right\} =\left\{ ({\bf \tilde{p}}, \tilde{\bm \theta }({\bf \tilde{p}})): {\bf \tilde{p}}\in A \right\} .
\end{equation}

We can observe that $\tilde{\bm \theta }({\bf p}(\bm \lambda_0 , \bm \eta_0))$ is an argmin of

$$ \psi (\bm \lambda , \bm \eta ):= D_{\phi }({\bf p}(\bm \lambda_0 , \bm \eta_0), {\bf p}(\bm \lambda , \bm \eta ))$$ because ${\bf p}(\bm \lambda_0 , \bm \eta_0)\in A$ and then

$$ {\bf F}({\bf p}(\bm \lambda_0 , \bm \eta_0), \tilde{\bm \theta }({\bf p}(\bm \lambda_0 , \bm \eta_0))) = {\partial D_{\phi }({\bf p}(\bm \lambda_0 , \bm \eta_0), {\bf p}(\tilde{\bm \theta }({\bf p}(\bm \lambda_0 , \bm \eta_0)))) \over \partial (\bm \lambda , \bm \eta )} = {\bf 0}.$$

On the other hand, applying \eqref{eq6},

$$ ({\bf p}(\bm \lambda_0 , \bm \eta_0), \tilde{\bm \theta }({\bf p}(\bm \lambda_0 , \bm \eta_0))) \in U,$$ and then ${\bf J}_{\bf F}$ is positive definite at $ ({\bf p}(\bm \lambda_0 , \bm \eta_0), \tilde{\bm \theta }({\bf p}(\bm \lambda_0 , \bm \eta_0))).$ Therefore,

$$ D_{\phi }({\bf p}(\bm \lambda_0 , \bm \eta_0), {\bf p}(\tilde{\bm \theta }({\bf p}(\bm \lambda_0 , \bm \eta_0)))) = \inf_{(\bm \lambda , \bm \eta )\in \Theta } D_{\phi }({\bf p}(\bm \lambda_0 , \bm \eta_0), {\bf p}(\bm \lambda , \bm \eta )),$$ and by the $\phi $-divergence properties $\tilde{\bm \theta }({\bf p}(\bm \lambda_0 , \bm \eta_0))= (\bm \lambda_0 , \bm \eta_0 )^T,$ and

$$ {\partial {\bf F}\over \partial {\bf p}(\bm \lambda_0 , \bm \eta_0) } + {\partial {\bf F} \over \partial (\bm \lambda_0 , \bm \eta_0)} {\partial (\bm \lambda_0 , \bm \eta_0 )\over \partial {\bf p}(\bm \lambda_0 , \bm \eta_0)} ={\bf 0}.$$

Further, we know that

$$ {\partial {\bf F} \over \partial (\bm \lambda_0 , \bm \eta_0 )} = \phi ''(1) {\bf A}(\bm \lambda_0 , \bm \eta_0)^T {\bf A}(\bm \lambda_0 , \bm \eta_0) $$ and we shall establish later that the $(t+u)\times 2^k$ matrix  $ {\partial {\bf F} \over \partial \bm \pi}$ is

\begin{equation}\label{eq7}
{\partial {\bf F} \over \partial {\bf p}(\bm \lambda_0 , \bm \eta_0)} = - \phi '' (1) {\bf A}(\bm \lambda_0 , \bm \eta_0)^T {\bf D}_{{\bf p}(\bm \lambda_0 , \bm \eta_0 )}^{-{1\over 2}}.
\end{equation}

Therefore, the $(t+u)\times 2^k$ matrix ${\partial (\bm \lambda_0 , \bm \eta_0 )\over \partial {\bf p}(\bm \lambda_0 , \bm \eta_0) }$ is

$$ {\partial (\bm \lambda_0 , \bm \eta_0 )\over \partial {\bf p}(\bm \lambda_0 , \bm \eta_0) } = ({\bf A}(\bm \lambda_0 , \bm \eta_0)^T{\bf A}(\bm \lambda_0 , \bm \eta_0))^{-1} {\bf A}(\bm \lambda_0 , \bm \eta_0)^T {\bf D}_{{\bf p}(\bm \lambda_0 , \bm \eta_0 )}^{-{1\over 2}}.$$

The Taylor expansion of the function $\tilde{\bm \theta }$ around ${\bf p}(\bm \lambda_0 , \bm \eta_0) $ yields

$$ \tilde{\bm \theta }({\bf \tilde{p}}) = \tilde{\bm \theta }({\bf p}(\bm \lambda_0 , \bm \eta_0)) + \left(  {\partial \tilde{\bm \theta }({\bf \tilde{p}}) \over {\bf \tilde{p}}}\right)_{{\bf \tilde{p}}=\bm \pi } ({\bf \tilde{p}} - {\bf p}(\bm \lambda_0 , \bm \eta_0)) + o(\| {\bf \tilde{p}} -{\bf p}(\bm \lambda_0 , \bm \eta_0) \| ).$$

As $\tilde{\bm \theta }({\bf p}(\bm \lambda_0 , \bm \eta_0)) = (\bm \lambda_0 , \bm \eta_0 )^T,$ we obtain from here

$$ \tilde{\bm \theta }({\bf \tilde{p}}) = (\bm \lambda_0 , \bm \eta_0 )^T+ ({\bf A}(\bm \lambda_0 , \bm \eta_0)^T{\bf A}(\bm \lambda_0 , \bm \eta_0))^{-1} {\bf A}(\bm \lambda_0 , \bm \eta_0)^T {\bf D}_{{\bf p}(\bm \lambda_0 , \bm \eta_0 }^{-{1\over 2}} ({\bf \tilde{p}} - {\bf p}(\bm \lambda_0 , \bm \eta_0)) + o(\| {\bf \tilde{p}} - {\bf p}(\bm \lambda_0 , \bm \eta_0) \| ).$$

We know that $ {\bf \hat{p}}{\overset{a.s.}{\longrightarrow }} {\bf p}(\bm \lambda_0 , \bm \eta_0 ),$ so that  ${\bf \hat{p}}\in A$ and, consequently, $ \tilde{\bm \theta }({\bf \hat{p}})$ is the unique solution of the system of equations

$$ {\partial  D_{\phi }( {\bf \hat{p}}, {\bf p} ( \tilde{\bm \theta }({\bf \hat{p}})))\over s_j} =0,\, j=1, ..., t+u,$$ and also $( {\bf \hat{p}}, \tilde{\bm \theta }({\bf \hat{p}}))\in U.$ Therefore, $\tilde{\bm \theta }({\bf \hat{p}})$ is the minimum $\phi $-divergence estimator, $\hat{\bm \theta}_{\phi}$, satisfying the relation

$$ \hat{\bm \theta}_{\phi} = (\bm \lambda_0 , \bm \eta_0 )^T+ ({\bf A}(\bm \lambda_0 , \bm \eta_0)^T{\bf A}(\bm \lambda_0 , \bm \eta_0))^{-1} {\bf A}(\bm \lambda_0 , \bm \eta_0)^T {\bf D}_{{\bf p}(\bm \lambda_0 , \bm \eta_0 )}^{-{1\over 2}} ({\bf \hat{p}} - {\bf p}(\bm \lambda_0 , \bm \eta_0 )) + o(\| {\bf \hat{p}} - {\bf p}(\bm \lambda_0 , \bm \eta_0 )\| ).$$

Finally, we are going to establish \eqref{eq7}. We compute the $(i,j)$-th element of the $(t+u)\times 2^k$ matrix ${\partial {\bf F}\over \partial {\bf p}(\bm \lambda_0 , \bm \eta_0) }.$

\begin{eqnarray*} {\partial \over \partial p_i} \left( {\partial D_{\phi } ({\bf \tilde{p}}, {\bf p}(\bm \lambda , \bm \eta ))\over \partial s_{j}} \right) & = & {\partial \over \partial p_i}  \left( \sum_{l=1}^{2^k} \left\{ \phi \left( {\tilde{p}_{l}\over p_l(\bm \lambda , \bm \eta )} \right) - \phi '\left( {\tilde{p}_{l}\over p_l(\bm \lambda , \bm \eta )} \right) {\tilde{p}_{l}\over p_l(\bm \lambda , \bm \eta )}\right\} {\partial p_l(\bm \lambda , \bm \eta )\over \partial s_j} \right) \\
& = & {1\over p_i(\bm \lambda , \bm \eta )} \left( -{p_i\over p_i (\bm \lambda , \bm \eta )} \phi ''\left( {p_i\over p_i (\bm \lambda , \bm \eta )}\right) \right) {\partial p_i(\bm \lambda , \bm \eta )\over \partial s_j}
\end{eqnarray*}
and for $(\pi_1, ..., \pi_{2^k}; \lambda_1^0, ..., \lambda_t^0; \eta_1^0, ..., \eta_u^0)$ we have

$$ {\partial \over \partial p_i} \left( {\partial D_{\phi } ({\bf \tilde{p}}, {\bf p}(\bm \lambda , \bm \eta ))\over \partial s_{j}} \right) = {1\over p_i (\bm \lambda_0 , \bm \eta_0 )} \phi ''\left( 1\right) {\partial p_i(\bm \lambda_0 , \bm \eta_0 )\over \partial s_j}.$$

Since ${\bf A}(\bm \lambda_0 , \bm \eta_0)={\bf D}_{{\bf p}(\bm \lambda_0 , \bm \eta_0 )}^{-{1\over 2}} {\bf J}(\bm \lambda_0 , \bm \eta_0),$ then \eqref{eq7} holds. \cqd

{\bf Proof of Theorem \ref{teo2}.}

Applying the previous theorem it holds

%
%
%

{\small $$ \sqrt{N}(\hat{\bm \theta}_{\phi} -(\bm \lambda_0 , \bm \eta_0)^T)  = \left( {\bf A}(\bm \lambda_0 , \bm \eta_0 )^T {\bf A}(\bm \lambda_0 , \bm \eta_0) \right)^{-1} {\bf A}(\bm \lambda_0 , \bm \eta_0) {\bf D}_{{\bf p}(\bm \lambda_0 , \bm \eta_0)}^{-{1\over 2}} \sqrt{N} (\hat{{\bf p}} - {\bf p}(\bm \lambda_0 , \bm \eta_0)) + \sqrt{N} \, \, \, o(\| \hat{{\bf p}} - {\bf p}(\bm \lambda_0 , \bm \eta_0) \| ).$$}

Note that

$$ \sqrt{N} \, \, \, o(\| \hat{{\bf p}} - {\bf p}(\bm \lambda_0 , \bm \eta_0) \| ) = o_p(1).$$

On the other hand, as $\hat{{\bf p}}$ is the sample proportion, we can apply the Central Limit Theorem to conclude

$$ \sqrt{N} (\hat{{\bf p}} - {\bf p}(\bm \lambda_0 , \bm \eta_0)) {\overset{L}{\longrightarrow }} {\cal N}({\bf 0} , \bm \Sigma_{{\bf p}(\bm \lambda_0, \bm \eta_0)} ), $$ where $\bm \Sigma_{{\bf p}(\bm \lambda_0, \bm \eta_0)} $ is given by

$$ \bm \Sigma_{{\bf p}(\bm \lambda_0, \bm \eta_0)} = {\bf D}_{{\bf p}(\bm \lambda_0 , \bm \eta_0)} - {\bf p}(\bm \lambda_0 , \bm \eta_0) {\bf p}(\bm \lambda_0 , \bm \eta_0)^T.$$

Therefore, it follows

$$ \sqrt{N}(\hat{\bm \theta}_{\phi} -(\bm \lambda_0 , \bm \eta_0 ))  {\overset{L}{\longrightarrow }} {\cal N}({\bf 0} , \bm \Sigma^*), $$ where $\bm \Sigma^* $ is given by


$$ \bm \Sigma^* = \left( {\bf A}(\bm \lambda_0 , \bm \eta_0 )^T {\bf A}(\bm \lambda_0 , \bm \eta_0) \right)^{-1} - {\bf B} {\bf B}^T$$
with ${\bf B}:=\left( {\bf A}(\bm \lambda_0 , \bm \eta_0 )^T {\bf A}(\bm \lambda_0 , \bm \eta_0) \right)^{-1} {\bf A}(\bm \lambda_0 , \bm \eta_0)^T {\bf D}_{{\bf p}(\bm \lambda_0 , \bm \eta_0)}^{1\over 2}.$

It is not difficult to see that

$$ {\bf D}_{{\bf p}(\bm \lambda_0 , \bm \eta_0)}^{{1\over 2}} {\bf A}(\bm \lambda_0 , \bm \eta_0)=0,$$ whence ${\bf B}={\bf 0}$ and the result holds. \cqd

{\bf Proof of Theorem \ref{teo3}.}

Using Theorem \ref{teo2}, it suffices to apply the delta method. Then, we can conclude that

$$ \sqrt{N}({\bf p}( \hat{\bm \theta}_{\phi})- {\bf p} (\bm \lambda_0, \bm \eta_0 )) {\overset{L}{\longrightarrow }} {\cal N}({\bf 0} , \nabla {\bf p}(\bm \lambda_0 , \bm \eta_0)^T\left( {\bf A}(\bm \lambda_0, \bm \eta_0 )^t {\bf A}(\bm \lambda_0, \bm \eta_0 ) \right)^{-1} \nabla {\bf p}(\bm \lambda_0 , \bm \eta_0)).$$

Now, as $\nabla {\bf p}(\bm \lambda_0 , \bm \eta_0)= {\bf J}(\bm \lambda_0 , \bm \eta_0)$, the theorem is proved. \cqd

\section*{{\bf References}}
\noindent

- Abar, B. and Loken, E. (2010). Self-regulated learning and self-directed study in a pre-college
  sample. {\em Learning and Individual Differences}, 20:25--29.

- Berkson, J. (1980). Minimum chi-square, not maximum likelihood! {\em Annals of Statisitcs}, 8(3):482--485.

- Biemer, P. (2011). {\em Latent Class Analysis and Survey Error}. John Wiley and Sons.

- Caldwell, L., Bradley, S., and Coffman, D. (2009). A person-centered approach to individualizing a scool-based universal preventive intervention. {\em American Journal of Drug and Alcohol Abuse}, 35(4):214--219.

- Clogg, C. (1995). Latent class models: Recent developments and prospects for the
future. In Arminger, C. G. and Sobol, M., editors, {\em Handbook of statistical modeling for the social and behavioral sciences}, pages 311--352. Plenum, New York (USA).

- Coffman, D., Patrick, M., Polen, L., Rhoades, B., and Ventura, A. (2007). Why do high school seniors drink? Implication for a targeted approach to intervention. {\em Prevention Science}, 8:1--8.

- Coleman, J.S. (1964). {\em Introduction to Mathematical Sociology}. Free Press, New York (USA).

- Collins, L. and Lanza, S. (2010). {\em Latent class and latent transition analysis for the social, behavioral, and health sciences}. Wiley, New York (USA).

- Cressie, N. and Pardo, L. (2002). Phi-divergence statisitcs. In: Elshaarawi, A.H., Plegorich, W.W. editors. {\it Encyclopedia of environmetrics}, vol. 13. pp: 1551--1555, John Wiley and sons, New York.

- Cressie, N. and Read, T. R. C. (1984). Multinomial goodness-of-fit tests. {\em J.
Roy. Statist. Soc. Ser. B}, 8:440--464.

- Csisz\'ar, I. (1967). Information-type measures of difference of probability distributions
  and indirect observations. {\em Studia Scientiarum Mathematicarum Hungarica}, 2:299--318.

- Feldman, B., Masyn, K., and Conger, R. (2009). New approaches to studying behaviors: A comparison of methods for modelling longitudinal, categorical and adolescent drinking data. {\em Development Psycology}, 45(3):652--676.

- Formann, A. (1976). Sch\"atzung der Parameter in Lazarsfeld Latent-Class Analysis. In {\em Res. Bull.}, number 18. Institut f\"ur Psycologie der Universit\"at Wien. In German.

- Formann, A. (1977). Log-linear Latent Class Analyse. In {\em Res. Bull.}, number 20. Institut f\"ur Psycologie der Universit\"at Wien. In German.

- Formann, A. (1978). A note on parametric estimation for Lazarsfeld's latent class analysis. {\em Psychometrika}, 48:123--126.

- Formann, A. (1982). Linear logistic latent class analysis. {\em Biometrical Journal}, 24:171--190.

- Formann, A. (1985). Constrained latent class models: Theory and applications. {\em British Journal of Mathematics and Statistical Psicology}, 38:87--111.

- Formann, A. (1992). Linear logistic latent class analysis for polytomous data. {\em Journal of the Amearican Statistical Association}, 87:476--486.

- Gerber, M., Witterkind, A., Grote, G., and Staffelbach, B. (2009). Exploring types of career orientation: a latent class analysis approach. {\em Journal of Vocational Behavior}, 75:303--318.

- Gill, P. E. and Murray, W. (1979). Conjugate-gradient methods for large-scale nonlinear optimization.
{\em Technical Report SOL 79-15}. Department of Operations Research, Stanford University.

- Goodman, L. A. (1974). Exploratory latent structure analysis using Goth identifiable and unidentifiable models. {\em Biometrika}, 61:215--231.

- Hagenaars, J. A. and Cutcheon, A. L. M. (2002). {\em Applied Latent Class Analysis}. Cambridge University Press, Cambridge (UK).

- Hooke, R. and Jeeves, T. A. (1961). Direct Search Solution of Numerical and statistical Problems. {\em Journal of the Association for Computing Machinery}, 8:212--229.

- Langeheine, R. and Rost, J. (1988). {\em Latent Trait and Latent Class Models}. Plenum Press, New York (USA).

- Laska, M., Pash, K., Lust, K., Story, M., and Ehlinger, E. (2009). Latent class analysis of lifestyle characteristics and health risk behaviors among college youth. {\em Prevention Sciences}, 10:376--386.

- Lazarsfeld, P. and Henry, N. (1968). {\em Latent structure analysis}. Houghton-Mifflin, Boston (USA).

- Lazarsfeld, P. (1950). The logical and mathematical foundation of latent structure analysis. In {\em Studies in Social Psycology in World War II, vol. IV: Measurement and prediction}, pages 362--412. Princeton University Press.

- McHugh, R. (1956). Efficient estimation and local identification in Latent Class Analysis. {\em Psychometrika}, 21:331--347.

- Morales, D., Pardo, L., and Vajda, I. (1995). Asymptotic divergence of estimators of discrete distributions. {\em Jounal of Statistical Planning and Inference}, 48:347--369.

- Nylund, K., Bellmore, A., Nishina, A., and Grahan, S. (2007). Subtypes, severity and structural stability of peer victimization: What does latent class analysis say? {\em Child Prevention}, 78:1706--1722.

- Pardo, L. (2006). {\em Statistical Inference based on Divergence Measures}. Chapman \& Hall CRC.

- Powell M.(1970). A hybrid method for nonlinear algebraic equations. In Rabinowitz, P. editor. {\em Numerical Methods for
Nonlinear Algebraic Equations}. Gordon and Breach.

- Rost, J. and Langeheine, R. (1997). {\em Applications of Latent trait and Latent Class Models in the Social Sciences}. Waxmann, M\"unster (Germany).

%
\end{document}